\documentclass[epj]{svjour}
\usepackage{epsfig}
\newcommand{\de}{\mbox{d}}
\newcommand{\e}{\mbox{e}}
\newcommand{\aco}{\alpha}
\newcommand{\nuc}[2]{{$^{#1}${#2}}}
\newcommand{\frc}[2]{\mbox{$\frac{#1}{#2}$}}
\newcommand{\ve}[1]{\vec{#1}}

\begin{document}

\title{{The one-body and two-body density matrices
 of finite nuclei
with an appropriate treatment of the center-of-mass motion
}\thanks{Supported in part by
the University of Athens under grant
70/4/3309 and by
the Deutsche Forschungsgemeinschaft
within the SFB 634.  
}}

\titlerunning{Density matrices of finite nuclei with an appropriate treatment of CM motion} 

\author{A.~Shebeko\thanks{shebeko@kipt.kharkov.ua} \inst{1} 
      \and P.~Papakonstantinou\thanks{panagiota.papakonstantinou@physik.tu-darmstadt.de} \inst{2} 
      \and E.~Mavrommatis\thanks{emavrom@cc.uoa.gr} \inst{3} 
        } 
\institute{NCS ``Kharkov Institute of Physics and Technology",
Academicheskaya Str. 1,
61108  Kharkov, Ukraine  
\and 
Institute of Nuclear Physics, T.U.Darmstadt, Schlossgartenstr. 9,
D-64289 Darmstadt, Germany 
\and 
University of Athens, Physics Department,
Nuclear and Particle Physics Division,
Panepistimiopoli, Ilissia, GR-157 71 Athens, Greece 
}

%\offprints{} 
%\mail{} 

\abstract{
The one-body and two-body density matrices in coordinate space and their Fourier
transforms in momentum space are studied for a nucleus 
(a nonrelativistic, self-bound finite system). 
%that consists of $A$ nucleons (particles). 
Unlike the usual procedure, suitable for infinite or 
externally bound systems, they are determined 
as expectation values of appropriate 
%$A$-particle multiplicative operators,
intrinsic operators, 
dependent on the relative
coordinates and momenta (Jacobi variables) 
and acting on 
intrinsic wavefunctions of nuclear states. 
Thus, translational invariance (TI) is respected. 
When handling such intrinsic quantities, 
we use an algebraic technique based upon the
Cartesian representation, 
in which the coordinate and momentum operators are 
linear combinations of the creation and annihilation operators ${\hat{\vec{a}}}^{+} $
and  $\hat {\vec{a}}$ for oscillator quanta. % in the three space directions.
Each of the relevant multiplicative operators can then be reduced to the form:
one exponential of the set $\{{\hat {\vec{a}}}^{+} \}$  times
other exponential of the set $\{ \hat{\vec{a}} \}$. 
In the course of such a normal-ordering
procedure we offer a fresh look at the appearance of ``Tassie-Barker" factors,
%find certain generating functions for the density and momentum distributions
and point out other model-independent results. 
The intrinsic wavefunction of the nucleus in its
ground state is constructed from a 
nontranslationally-invariant (nTI) one via 
existing projection techniques. 
%the prescriptions
%after Ernst, Shakin and Thaler, 
%on the one hand, and after Peierls and Yoccoz, on the other
%hand. 
As an illustration, 
the one-body and two-body momentum distributions 
(MDs) for
the \nuc{4}{He} nucleus are calculated 
with the Slater
determinant 
of the harmonic-oscillator model as the trial, nTI wavefunction.
We find that the TI introduces important effects in the MDs. 
}


\maketitle 

\section{Introduction}

The last few years the interest in the study of nuclei from both 
experimental and theoretical point of view has shifted from the 
investigation of one-body quantities 
(e.g., the elastic form factor $F(\vec{q})$ and the momentum distribution 
$\eta (\vec{p})$) 
towards the 
investigation of two-body quantities, with the aim of revealing more 
direct information on the dynamical correlations between the nucleons 
(short-range (SRC) and tensor). The two-body quantities are connected to 
the two-body density matrix (2DM) in coordinate or momentum space as are 
the one-body quantities connected to the one-body density matrix (1DM). 
The 2DM, besides being interesting in itself, allows the calculation of the 
expectation value of any two-body operator \cite{Low55}. 
In addition to the 2DM 
$\rho^{[2]}(\vec{r}_1,\vec{r}_2;\vec{r}_{1'},\vec{r}_{2'})$, 
we will 
also consider the two-body momentum distribution (TBMD) 
$\eta^{[2]}(\vec{p},\vec{k})$, 
which is the Fourier 
transform of the 
$\rho^{[2]}(\vec{r}_1,\vec{r}_2;\vec{r}_{1'},\vec{r}_{2'})$ 
in the variables 
$\vec{r}_1-\vec{r}_2$  
and  
$\vec{r}_{1'}-\vec{r}_{2'}$  
and is connected to the 
two-nucleon spectral function 
$S(\vec{p},\vec{k};E)$ 
via integration with respect to the energy $E$.

A prominent role towards the experimental investigation of the 
2DM and related quantities is played by the study of the electromagnetically 
induced 2-nucleon emission 
%((γ, ΝΝ), e, eNN) 
($\gamma$,NN), (e,e'NN) 
which can be carried out with 
high accuracy in photon facilities (Elsa, MAMI) and electron accelerators with 
high energy, 100\%~duty-cycle beams [Jefferson Lab, MAMI]. 
Past, present and near future experiments provide these 
useful data \cite{BF90,BGG03,Wal03,Nig04,Sta04}. 
Theoretical methods to analyze the mechanisms of these reactions and to 
calculate the relevant nuclear two-body properties are under 
continuous development. 
In particular, for the case of finite nuclei the 
generalized momentum distribution 
$\eta (\vec{p},\vec{Q})$  \cite{PMK2000,MPM2099}, 
the two-body momentum distribution  
$\eta^{[2]}(\vec{p},\vec{k})$ 
and other two-body distributions 
have been studied for $Z = N$,  
$\ell -$closed nuclei, 
as well as the two-body density matrix for the nuclei 
\nuc{4}{He} \cite{OrS95}, 
\nuc{16}{O} and \nuc{40}{Ca} \cite{DKA2000} 
and the two-nucleon spectral function  
$S(\vec{p},\vec{k};E)$ 
for the nucleus \nuc{16}{O} 
(see  \cite{PMK2003} and refs. therein).

One of the theoretical issues still under discussion is the proper 
consideration of the requirement of translational invariance and 
therefore the conservation of the total momentum of the system. 
The wavefunctions which have been used in the 
independent-particle shell model and in theories 
which take also dynamical correlations into account (e.g. Brueckner-Hartree-Fock, 
Variational Monte Carlo) satisfy the Pauli principle but not the 
translational invariance. As a consequence, they contain “spurious” 
components which result from the motion of the Center of Mass (CM) 
in a non free state. 
Effects from these (also known as CM correlations) 
are found in the calculation of almost every observable and make 
impossible the extraction of information for the intrinsic properties of 
nuclei directly from the experimental data. In addition, there is 
ambiguity in the proper definition of translationally invariant 
operators which correspond to the different physical quantities.

Many efforts have been made towards the solution of the CM %Center of Mass 
problem. In some of them the treatment of the CM motion is built right 
into the theory one is using \cite{ES55,DdF74,dF80}. 
In the majority of the efforts 
%though 
the restoration of translational invariance is attempted after 
the wavefunctions have been developed. One such approach consisted of 
adding intuitively the three extra degrees of freedom 
of the CM to the $3A$ internal 
coordinates. Work along these lines using configurations from the 
harmonic oscillator model (HOM) has been carried out 
by Tassie and Barker \cite{TB58} 
and others. Most of the other approaches use projection techniques to 
define suitable intrinsic wavefunctions with coordinates 
referred to 
the CM. 
%The way, however, with which one projects, is crucial. 
The pioneering works along this direction were made in the 50s 
by 
Gartenhaus and Schwartz (GS) \cite{GaS1957} 
and Peierls and J. Yoccoz (PY) \cite{PeY1957},  
with violations of the Gallilei invariance (GI), 
followed by 
Ernst, Shakin and Thaler (EST) \cite{EST1973,EST1973b},  
with their critique on the GS transformation, 
Vincent \cite{Vin1973}, 
Shebeko et al \cite{ShG1974,DOS1975} and 
others in the 70s \cite{DdF74,Fri1971}. 
%others in the 70s \cite{dF80,Fri1971}. 
Projection techniques have been proposed by 
Schmid and Gr{\"u}mmer \cite{ScG1990a,ScG1990} 
at the beginning of the 90’s and rather recently by 
Schmid and collaborators using harmonic-oscillator \cite{Schmid} and spherical 
Hartree-Fock configurations \cite{RGS2004}. 
We should add that Mihaila and Heisenberg have worked out the problem 
of CM corrections by expanding them as many-body operators \cite{MiH99}. 
It seems that if the wavefunction is very nearly factorable into a 
center-of-mass and an intrinsic component all the approaches to 
treat the CM problem are equivalent, provided that the translationally 
invariant operators are used. Also, all the approaches can be 
carried out rather simply for the independent particle shell model with 
harmonic oscillator potential.

The consideration of CM effects with one or more of the above 
mentioned methods have mostly addressed the light nucleus \nuc{4}{He} and 
single-particle quantities: the kinetic energy, the single - particle energies, 
the one-body density matrix, the matter and charge density, the 
elastic form factor, the dynamic structure factor, the momentum 
distribution and occupation probabilities, the one-body spectral function, 
the single-particle overlap function etc. There are few calculations 
for other light-medium nuclei such as \nuc{12}{C}, \nuc{15}{N}, \nuc{16}{O}, \nuc{40}{Ca}. 
The consideration of CM correlations in two-body quantities has been 
limited so far, to the best of our knowledge to the potential energy. 
%The different methods have yielded different results for the same quantity. 
%They are arguments in favor of one or the other method.

In this paper, we will start with the evaluation of  
the one-body density matrix in coordinate space and 
related one-body momentum distribution (OBMD). 
As mentioned above, CM effects have been 
considered in the calculation of 1DM and OBMD before. 
By starting with such evaluation we want to present our 
method and technicalities. We then proceed to evaluate 
the two-body density matrix and two-body momentum distribution. 
We will take into account the CM correlations 
by using the EST prescription or fixed-CM approximation to 
construct the intrinsic wavefunction. We also use a specific 
prescription for defining the corresponding intrinsic operators. 
As mentioned before, 
the EST method has been introduced in refs.~\cite{EST1973,EST1973b}. 
Subsequently, it has been used in the calculation of the 
elastic and dynamic form factor and the momentum distribution of 
\nuc{4}{He} \cite{DOS1975,GKS2002}. 
The EST intrinsic many-body wavefunction is constructed from a 
translationally non-invariant one by projecting onto an 
eigenstate of total momentum using a non unitary operator which 
fixes the CM coordinate $R$ to be equal to zero. 
This 
transformation has certain advantages compared to GS 
transformation \cite{EST1973b} (in particular, it ensures a correct 
behavior under Galilean transformation). 
It has turned out \cite{EST1973b} (see also \cite{DOS1975})  
that the GS transformation can be reduced to the EST projection 
procedure.  
As for the relevant one- and two-body intrinsic operators, 
they are defined %in the usual way 
by %simply 
replacing the 
coordinates and momenta by relative ones (Jacobi variables). 
Unlike the definitions of the overlap integrals 
with intrinsic wavefunctions used in ref.~\cite{VNW1998}, 
we are dealing with the expectation 
values of the intrinsic operators as they occur 
under the treatment of the aforementioned quantities. 
Subsequently, we present the way for calculating the matrix 
elements defined by the above intrinsic wavefunctions and 
operators using an algebraic technique introduced 
in ref.~\cite{ShG1974} 
and based upon the Cartesian representation of the coordinate and 
momentum operators in terms of linear combinations of the 
creation and annihilation vector operators 
(${\hat{\vec{a}}}^{+} $
and  $\hat{\vec{a}}$ respectively) 
for 
oscillator quanta in the three different space directions. 
With this technique one avoids to deal with difficult multiple 
integrals and therefore it seems to be the technique of choice 
in the case of systems with large number of bodies. 
The Cartesian representation is particularly convenient 
in the case of wavefunctions constructed with Slater determinants. 
The application of the above to the evaluation of  the OBMD leads to the 
derivation of the Tassie-Barker factor \cite{TB58} 
($\exp{r_0^2q^2/4A}$, 
$r_0$ oscillator length 
parameter entering into the definition of   
${\hat{\vec{a}}}^{+} $
and  $\hat {\vec{a}}$ - see sect.~2) 
in a model-independent way. In addition, the evaluation of both OBMD and 
TBMD leads to other model-independent results. 

Next, the intrinsic OBMD and TBMD are evaluated in the 
independent particle shell model with harmonic oscillator 
wavefunctions. Such a model leads to compact analytical expressions. 
Moreover, it is expected that the results derived with its use will be 
close to the ones that will be obtained with the use of more realistic 
single-particle wavefunctions (Woods-Saxon or Hartree-Fock), 
since the quantities under study are not defined in terms of the 
asymptotic behavior of the wavefunction for large $r$. 
The latter is wrong in the case of the wavefunction in the HOM.

In this work, the above evaluation is carried out for the 
intrinsic OBMD and TBMD of the nucleus \nuc{4}{He}. 
It is expected that the CM effects are more pronounced for 
light nuclei. In addition, due to its high central density 
(almost 3 times nuclear matter density) the nucleus \nuc{4}{He} is a 
particularly appropriate system to search for the origin of SRC. 
The evaluation of the intrinsic OBMD of \nuc{4}{He} has appeared 
before in the case of HOM \cite{DOS1975} as well as other 
single-particle models \cite{GKS2002}. 
%The experimental study of the 
%TBMD of \nuc{4}{He} at Jefferson Lab follows the one in the case of 
%\nuc{3}{He}. 
The (not intrinsic) TBMD of \nuc{4}{He} has been studied in 
refs.~\cite{PMK2003,Pap2004} by including in the HOM 
Jastrow-type correlations via the lowest term of the so-called 
low-order approximation \cite{DSB1982}, but ignoring CM correlations. 
A comparison of the present results with those of refs \cite{PMK2003,Pap2004} 
will reveal the relative importance of the CM and SRC correlations 
in the same nucleus. The effect of 
the different correlations is estimated by 
introducing the quantity 
$\eta^{[2]}(\vec{p},\vec{k})/\eta(\vec{p})\eta(\vec{k})$. 
We find that the CM correction 
reduces the width of OBMD and TBMD and in the case of the TBMD 
introduces a dependence on the angle between $\vec{p}$ and $\vec{k}$, 
a shift of its 
peak in favor of opposite momenta and significant 
deviations for large values of $p$ and $k$ 
and angles close to 180$^{\circ}$. 
This last effect is also found when SRC are considered.
Up to now, the TBMD of 
\nuc{3}{He} has been experimentally studied 
at Jefferson Lab \cite{Nig04}. 
%\nuc{4}{He} has not been experimentally studied. 
%One can use the study of the TBMD of \nuc{3}{He} 
%at Jefferson Lab \cite{Nig04} to extract some useful information. 

The paper is organized as follows. 
In Section 2, we present the general formalism of 
constructing appropriate wavefunctions that respect 
translational invariance using the Ernst-Shakin-Thaler 
prescription 
and describe the evaluation of matrix elements 
of intrinsic operators usung the Cartesian representation. 
In Section 3 the definitions of the relevant 
operators for the intrinsic quantities under study, namely 
the one-body quantities 1DM, form factor and OBMD and the 
two-body quantities 2DM and TBMD, are introduced in terms of the 
relative coordinates and momenta (Jacobi variables). 
In Section 4 the above quantities are evaluated using the 
Cartesian representation. In Section 5 by considering 
specifically the independent-particle shell model with 
harmonic oscillator wavefunctions, results are derived and 
discussed for the OBMD and TBMD of the nucleus \nuc{4}{He}. 
Finally, in Section 6 a summary of the results and hints 
for possible further work are given.

\section{Constructing intrinsic wavefunctions
and matrix elements. The Cartesian representation
}

Let us consider a nonrelativistic system
composed of $A$ particles (nucleons).
The coordinate (momentum) vector of the $\alpha -$th particle
will be denoted by $\vec{r}_{\alpha}$ ($\vec{p}_{\alpha}$).
Occasionally, we will use the generic symbol
$\alpha$,
which may include
spin and/or isospin degrees of freedom,
but in most cases we will suppress these degrees of freedom
for the sake of simplicity.

In principle,
the eigenvectors
of the total Hamiltonian $\hat{H}$ of the system $| \Psi_{\ve{P}} \rangle $,
which belong to
the eigenvalue $\ve{P}$ of the
total momentum operator $\hat {\ve{P}}$, can be written
as the product
\begin{equation}
| \Psi_{\ve{P}} \rangle = | \ve{P} )\ | \Psi_{\rm int} \rangle .
\end{equation}
Following ref.~\cite{EST1973},
the bracket $|\,\,) $ is used to represent a
vector in the space of the center-of-mass coordinates, so that
$\hat {\ve{P}} | \ve{P} ) = \ve{P}  | \ve{P} ) $.
A ket~(bra) with an index
$|\cdots\rangle_{\alpha}$~($_{\alpha}\langle \cdots|$)
will refer to the state of the $\alpha-$th particle.
The intrinsic wavefunction 
$\Psi_{\rm int}$ depends upon the $A-1$ independent
intrinsic variables.
These may be expressed in terms of the
Jacobi coordinates
\begin{equation}
\label{e:ja}
{{\ve{\xi}}}_{\alpha} = {{\ve{r}} }_{\alpha +1} -
\frac{1}{{\alpha}}\sum_{\beta =1}^{\alpha} {{\ve{r}}}_{\beta}
\qquad (\alpha = 1,2, \ldots ,A-1) \,
\end{equation}
or
the corresponding
canonically conjugate momenta
\begin{equation}
{{\ve{\eta}}}_{\alpha} = \frac{1}{\alpha +1} (\alpha {{\ve{p}}}_{\alpha +1} -
\sum_{\beta =1}^{\alpha}
{{\ve{p}}}_{\beta} )\qquad (\alpha = 1,2, \ldots ,A-1)
 .
\end{equation}
The wavefunction $\Psi_{ \ve{P}}  (\ve{r}_1 , \ve{r}_2 , ... , \ve{r}_A) $
in the coordinate representation satisfies the requirement of
translational invariance,
\begin{equation}
\label{Etrinv}
\Psi_{\ve{P}}  (\ve{r}_1 + \ve{a}, \ve{r}_2 + \ve{a}, \ldots
, \ve{r}_A + \ve{a} ) = \exp ({\rm i} \ve{P}\cdot\ve{a})\Psi_{ \ve{P}}
(\ve{r}_1 , \ve{r}_2 , ... , \ve{r}_A)
 ,
\end{equation}
for any arbitrary displacement $\ve{a}$.

When describing scattering processes,
it is convenient to consider
the initial target state $| 0 \rangle $ as a
%superposition,
$\ve{P}$--packet,
\begin{equation}
\label{e:zero}
| 0 \rangle = \int | \Psi_{\ve{P}}
\rangle {\rm d} \ve{P} \langle \Psi_{ \ve{P}}
| 0 \rangle \equiv \int c(\ve{P}) |
\Psi_{\ve{P}} \rangle \, {\rm d}^3 P
\end{equation}
(see also \cite{GW64}, Ch.~XI),
with the normalization condition
\begin{equation}
\label{e:anor}
\langle 0 \mid 0 \rangle = \int | c(\ve{P}) |^2 \, {\rm d}^3 P = 1
.
\end{equation}
Being the exact $\hat{H}$--eigenvectors,
the states $| \Psi_{\ve{P}} \rangle $ belong
simultaneously to
the set of eigenvectors
of the total momentum operator $\hat {\ve{P}}$
with eigenvalues $\ve{P}$
close to a
given value $\ve{P}_t$, e.g., $\ve{P}_t = 0 $.
The final state of the recoiling nucleus
is written in the form
$|\Psi_{\vec{P'}}\rangle$.
Evidently,
the wavepacket $|0\rangle$
is not translationally invariant.
However, this shortcoming can be corrected
by letting the width of the packet $\Delta$ go to zero at the end of the calculations,
i.e., assuming that
\begin{equation}
\label{e:lim}
\lim_{\Delta \rightarrow 0} \int | c(\ve{P}) |^2
g(\ve{P}) {\rm d}^3 P = \int \delta (\ve{P} - \ve{P}_t) g(\ve{P})
{\rm d} \ve{P} = g (\ve{P}_t)
\end{equation}
for an arbitrary function $ g(\ve{P})$.

This prescription has a
transparent physical meaning being adequate
to many scattering situations. With its aid one can express the corresponding cross sections
in terms of intrinsic quantities.
Let us consider, for instance, the elastic scattering of a
particle from
the nucleus.
In the Plane Wave Impulse Approximation (PWIA)
and neglecting the Fermi-motion effects,
the cross section of interest
can be represented in the form
\begin{equation}
\label{e:sig1}
\sigma (\theta) = \lim_{\Delta \rightarrow 0}
\int K(\ve{P'})\, {\rm d}^3 P'
|\langle \Psi_{ \ve{P'}} | \exp [{\rm i}\ve{q}
\cdot {\hat{\ve{r}}}_A ]
| 0 \rangle |^2 \ ,
\end{equation}
where
$\vec{q}$ is the momentum transfer,
$\theta$ the scattering angle
and $K(\ve{P'}) $ the corresponding kinematical
factor.
By substituting $|0\rangle $ from eq.~(\ref{e:zero}), using eq.~(1)
and writing $\hat{\vec{r}}_A =
(\hat{\vec{r}}_A - \hat{\vec{R}})
+ \hat{\vec{R}}$, we evaluate
$\langle \Psi_{ \ve{P'}} |
\exp [{\rm i}\ve{q} \cdot {\hat{\ve{r}}}_A ] | 0 \rangle
=\int
\de^3P c(\ve{P})
(\vec{P'}|
\exp ({\rm i}\vec{q}\cdot\hat{\vec{R}})  |\vec{P})
\langle \Psi_{\rm int} |
 \exp{[{\rm i}\vec{q}
\cdot
(\hat{\vec{r}}_A - \hat{\vec{R}})]}
|\Psi_{\rm int}\rangle =
\int \de^3P c(\ve{P})
\delta (\ve{P'} - \ve{P} - \ve{q} )
F_{\rm int} (\ve{q})
$,
that leads to
\[
\sigma (\theta) = K(\ve{P}_t + \ve{q}) \Bigl| F_{\rm int} (\ve{q})\Bigl|^2 ,
\]
where the elastic\,(intrinsic) form factor (FF) 
 $F_{\rm int} (\vec{q})$ is determined by
\begin{equation}
\label{e:ff1}
F_{\rm int} (\vec{q}) = \langle \Psi_{\rm int}
| \exp [{\rm i}\ve{q}\cdot
 (\hat{\ve{r}}_A - \hat {\ve{R} } ) ] |
\Psi_{\rm int} \rangle
\equiv \langle \Psi_{\rm int} |
\hat{F}_{\rm int}(\vec{q})
| \Psi_{\rm int} \rangle .
\end{equation}
Use has been made, on the one hand, of the fact
that intrinsic and CM operators commute with each other
and, on the other hand, of the condition (\ref{e:lim}).
The $F_{\mathrm{int}}(\vec{q})$ 
is typical of the quantities of interest,
viz., it is the expectation value of an operator
that depends on intrinsic coordinates, namely
$\hat{\ve{r}}_A - \hat {\ve{R} } $.

As we have mentioned in the Introduction, our  aim is to calculate expectation values
of one- and two-body operators in
the nuclear ground state\,(g.s.)
taking into account the requirement for TI.
We are interested,
in particular,
in intrinsic quantities which  appear in
analytical expressions describing various scattering
cross sections off a nucleus (in general,
a finite system).
Such quantities include,
besides the elastic
FF $F(\vec{q})$,
the particle density $\rho (\vec{r})$,
the dynamical FF
$S(\vec{q}, \omega)$,
%entering the description of the inclusive $(e,e')$ scattering,
the OBMD
$\eta (\vec{p})$
which is often associated
with the one-body spectral function $P(\vec{p}, E)$,
and the TBMD $\eta^{[2]}(\vec{p},\vec{k})$
that is related to the two-body spectral function
$ S(\vec{p},\vec{k}; E) $.

%that may be used,
%under some assumptions, for the description of
%the $(e,e'2N)$ triple-coincidence cross sections.

At the initial stage of the calculations, the nuclear g.s. 
is represented by a  ${\vec P}$ -- packet as introduced above.
Then the main task is to construct
the TI wavefunctions
$|\Psi_{\vec{P}}\rangle$ in a tractable manner,
so that the CM-motion separation can be achieved.
It is important to properly define
the quantities of interest in terms of intrinsic
coordinates, as was already done for the FF
in eq.~(\ref{e:ff1}). We will tackle
this issue in the next section.

First of all, let us consider a Slater determinant
\begin{equation}
\label{Eslat}
|{\rm Det}\rangle = \frac{1}{\sqrt{A!}} \sum_{\it \hat{\mathcal{P}} \in S_A}
\epsilon_{ \mathcal{P}}
\hat{\mathcal{P}} \{
|\phi_{p_1}(1)\rangle
|\phi_{p_2}(2)\rangle
\cdots
|\phi_{p_A}(A)\rangle
\} \,
\end{equation}
as the total  wavefunction
$|\Psi_0\rangle$
for
an approximate and convenient
description of the nuclear g.s., in the framework
of the independent-particle
model or the Hartree-Fock approach.
In eq.~(\ref{Eslat}), $\epsilon_{\mathcal{P}}$ 
is the parity factor
for the permutation $\hat{\mathcal{P}}$, $\phi_{p_{\alpha}}$
are the occupied single-particle orbitals and
the summation
runs over all permutations of
the symmetric group $S_A$.

The wavefunction (\ref{Eslat})
 exemplifies wavefunctions which do not possess the property
of TI, eq.~(\ref{Etrinv}).
Obviously, any wavefunction that is constructed by
acting on $|{\rm Det}\rangle$ with a two- or a three- 
body correlation operator 
(e.g. a Jastrow correlation factor)
will not be translationally invariant either.
There are different ways to restore TI
if one starts with a ``bad" wavefunction $|\Psi\rangle$
such as $|{\rm Det}\rangle$
%\cite{PeY1957,GaS1957,EST1973,ScG1990,Schmid}. %Sch2001,Sch2002a,Sch2002b}.
\cite{PeY1957,EST1973,ScG1990,Schmid}. %Sch2001,Sch2002a,Sch2002b}.
We shall employ the so-called
``fixed-CM approximation",
or EST prescription \cite{EST1973}. 
{However, 
other projection recipes can be applied 
without essential changes  - we will come
back to this point at the end of the Sec.~4.} 
Within the EST approach, 
the approximate complete wavefunction is determined by
\begin{equation}
| \Psi_{\vec{P}}^{\rm EST} \rangle = | \ve{P} ) | \Psi_{\rm int}^{\rm EST} \rangle
\end{equation}
and the intrinsic unit-normalized wavefunction
is given by
\begin{equation}
\label{e:esti}
|\Psi_{\rm int}^{\rm EST} \rangle =
%(2\pi )^{3/2}
{(\vec{R}=0 |\Psi\rangle }  /
{ [{\langle \Psi | \delta (\hat{\vec{R}}) |\Psi\rangle }]^{1/2} }
 ,
\end{equation}
where $ ( \vec{R}=0 | $ is the eigenvector of  the CM operator,
$\hat{\ve{R}} = A^{-1} \sum_{\alpha =1}^A \hat{\ve{r}}_{\alpha}
$.
We have used the relation
$|\vec{R}=\vec{X})(\vec{R}=\vec{X}|=\delta (\hat{\vec{R}}-\vec{X})$.
For simplicity,
we confine ourselves to the case of identical particles
with mass $m$.
The complete EST wavefunction can now be represented as
\begin{equation}
| \Psi_{\vec{P}}^{\rm EST} \rangle = {U_{\vec{P}} |\Psi\rangle }  /
{ [{\langle \Psi | (2\pi )^3 \delta (\hat{\vec{R}}) | \Psi\rangle }]^{1/2} }
 , \end{equation}
where, following \cite{EST1973}, we have introduced the projection operator
($U_{\vec{P}}^2 = U_{\vec{P}} $)
\begin{equation}
 U_{\vec{P}} \equiv (2\pi )^{3/2} |\vec{P}) (\vec{R}=0|  .
\end{equation}
Given an operator $\hat{A}$,
its matrix elements with the TI symmetry can be written in the form
\begin{equation}
\label{e:estme}
\langle \Psi'_{\vec{P'}}|\hat{A}|\Psi_{\vec{P}}\rangle = \frac{\langle \Psi'|
 U^+_{\vec{P'}} \hat{A}
 U_{\vec{P}}
 | \Psi \rangle }{
  [ {\langle\Psi '| (2\pi )^3 \delta (\hat{\vec{R}}) |\Psi '\rangle
        \langle\Psi  | (2\pi )^3 \delta (\hat{\vec{R}}) |\Psi  \rangle }]^{1/2} } .
\end{equation}
Its expectation value in the g.s.
$$
|0\rangle = \int \de^3P c(\vec{P})
 {U_{\vec{P}} |\Psi_0\rangle}   /
{ [{\langle \Psi_0 | (2\pi )^3 \delta (\hat{\vec{R}}) | \Psi_0\rangle }]^{1/2}}
$$
is expressed in terms of the expectation value of the operator $  U^+_{\vec{P'}} \hat{A}
 U_{\vec{P}} $,
\begin{equation}
\langle 0 | \hat{A} |0\rangle  = \int\de^3 P \int \de^3 P' c^{\ast}(\vec{P'}) c(\vec{P})
 \frac{\langle \Psi_0|
 U^+_{\vec{P'}} \hat{A}
 U_{\vec{P}}
 | \Psi_0 \rangle }{
  \langle\Psi_0 |(2\pi )^3 \delta (\hat{\vec{R}}) |\Psi_0 \rangle } .
\end{equation}
%In other words, one has to deal with the operator $  U^+_{\vec{P'}} \hat{A} U_{\vec{P}} $.
%
In addition, if $\hat{A}$
is an intrinsic operator $\hat{A}_{\rm int}$,
acting only
on the space of intrinsic variables,
we find
\begin{equation}
\label{e:estoin1}
A_{\rm EST} \equiv
\langle 0 | \hat{A}_{\rm int} |0\rangle = {\langle \Psi_0|
 \hat{A}_{\rm EST}
 | \Psi_0 \rangle } / {
  \langle\Psi_0 | \delta (\hat{\vec{R}})
|\Psi_0 \rangle
}
\end{equation}
\begin{equation}
\label{e:estoin}
\hat{A}_{\rm EST} =
%(2\pi)^3
| \vec{R}=0 ) \hat{A}_{\rm int} ( \vec{R}=0 | =
%(2\pi)^3
\delta (\hat{\vec{R}}) \hat{A}_{\rm int} =
%(2\pi)^3
\hat{A}_{\rm int} \delta (\hat{\vec{R}}) .
\end{equation}
When deriving eq.~(\ref{e:estoin1} ) we have employed the relation
$(\vec{P}|\vec{P'}) = \delta (\vec{P}-\vec{P'})$ and equation (\ref{e:anor}).

It has been shown \cite{DOS1975}
that the calculation of
expectation values of  many-body operators
like $\hat{A}_{\rm EST}$ can be
substantially simplified using the Cartesian representation.
In this representation
the coordinate\,(momentum) operator
$\hat{\vec{r}}_{\alpha}$
($\hat{\vec{p}}_{\alpha}$) of the $\alpha$-th particle
is expressed through the Cartesian
creation and annihilation operators
${\hat{\vec{a}}}^{+} $ and $\hat{\vec{a}}$ ,
\begin{equation}
\label{e:cao}
\hat{\vec{r}} = \frac{r_0}{\sqrt{2}} ({\hat{\vec{a}}}^{+}
+ {\hat{\vec{a}}})
      \hspace{4mm}
\hat{\vec{p}} = {\rm i} \frac{p_0}{\sqrt{2}}
({\hat{\vec{a}}}^{+}
- {\hat{\vec{a}}})
      \hspace{4mm}
r_0p_0 = 1  , 
\end{equation}
obeying the commutation relations
\begin{equation}
\label{e:com1}
 [ \hat{a}^+_l,\hat{a}^+_j ] =  [ \hat{a}_l,\hat{a}_j ] = 0
      \hspace{5mm} , 
      \hspace{5mm}
 [ \hat{a}_l,\hat{a}^+_j ] = \delta_{lj} , 
      \hspace{1cm}
\end{equation}
which are the stepping stones in what follows.
The indices $l,j=1,2,3$ label the three Cartesian axes
$x,y,z$.

As the ``length" parameter $r_0$
one can choose the
oscillator parameter of a suitable
harmonic oscillator basis in which the nuclear wavefunction
is expanded.
Its basis vectors
$|n_x\,n_y\,n_z\rangle_1 \otimes \ldots \otimes
|n_x\,n_y\,n_z\rangle_A $,
where the quantum numbers $n_x,\,n_y,\,n_z$
take the values
$0,1,\ldots ,$
 are composed of the single--particle states
\begin{equation}
\label{e:bas}
|n_x\,n_y\,n_z\rangle = \left[n_x !\,n_y !\,n_z !
\right]^{-\frac12}
\left[ \hat{a}_1^{+} \right]^{n_x}
\,\left[ \hat{a}_2^{+} \right]^{n_y} \,
\left[ \hat{a}_3^{+} \right]^{n_z} | 0\,0\,0 \rangle  \,
 ,
\end{equation}
which are the eigenstates of the Hamiltonian
$\hat{H}_{\rm osc} =
\omega ( \hat{\vec{a}}^{+} \cdot \hat{\vec{a}} + \frac32 ) $,
\[
\hat{H}_{\rm osc} |n_x\,n_y\,n_z\rangle =
( n_x + n_y + n_z + \mbox{$\frac32$} )\, \omega \,
|n_x\,n_y\,n_z\rangle \ ,
\]
where $\omega$ is
the oscillation frequency
along the three axes $x,y$ and $z$.
{We use the system of units with $\hbar=c=1$}.
The single--particle wavefunction in coordinate representation
is written
$$
\langle \ve{r} \mid n_x\,n_y\,n_z\rangle =
\psi_{n_x} (x) \psi_{n_y} (y) \psi_{n_z} (z) \,
 ,
$$
where \cite{BoM69,NS69}
$$
\psi_{n} (s) = \left[ \sqrt{\pi} 2^n n! r_0
\right]^{- \frac12}
H_n ( {s / r_0} ) \exp (- {s^2 / 2 r_0^2} )
$$
and $H_n(x)$
is a Hermite polynomial.
By definition, the oscillator parameter equals
$r_0 = [m \omega]^{-\frac12} $.
%Recall also that the lowest-energy state $| 0\,0\,0 \rangle$
%with $n_x = n_y = n_z = 0 $ is destroyed
%by the operator $\hat{\ve{a}} $ , i.e.,
%$\hat{\ve{a}}| 0\,0\,0 \rangle = 0 $ .

The general idea in subsequent manipulations is
to bring a given operator
into a form with normal ordering,
in which the destruction operators
$\hat{\vec{a}}$
are to the right with respect to the creation operators
$\hat{\vec{a}}^{+}$ (see Secs.~4,~5).
For this purpose, we will also make use of the operator identity
\begin{equation}
\label{e:exp}
e^{\hat{A}+\hat{B}} 
= e^{\hat{A}}
e^{\hat{B}}
e^{-\frac{1}{2} \hat{C} } 
= e^{\hat{B}}
e^{\hat{A}}
e^{\frac{1}{2} \hat{C} }
 ,
\end{equation}
which is valid for arbitrary operators $\hat{A}$ and $\hat{B}$
if the operator $ \hat{C} = [\hat{A},\hat{B}] $
commutes with each of them.
In particular,
\begin{equation}
\label{e:vexp}
e^{\vec{x}\cdot\hat{\vec{A}}+\vec{y}\cdot\hat{\vec{B}}} 
= e^{\vec{x}\cdot\hat{\vec{A}}}
e^{\vec{y}\cdot\hat{\vec{B}}}
e^{-\frac{1}{2}\vec{x}\cdot\vec{y}C } 
= e^{\vec{y}\cdot\hat{\vec{B}}}
e^{\vec{x}\cdot\hat{\vec{A}}}
e^{ \frac{1}{2}\vec{x}\cdot\vec{y}C }
 ,
\end{equation}
if $[\hat{A}_l,\hat{B}_j]=C\delta_{lj} $ for $(l,j = 1,2,3)$
and $C$ is a $c-$number.
% The indices $l,j$ label the three Cartesian axes.

\section{The intrinsic  density matrices and related quantities}

In the preceding discussion it is implied that the
operators of interest have been expressed
in terms of the relevant coordinates,
e.g., intrinsic ones.
It is not always straightforward how to do this.
Here we refer mainly to the definitions
of $n-$body density matrices (nDM's).
For instance, it is a
common practice \cite{MYS67,AHP88,AHP93},
to write the 1DM in coordinate representation
as the expectation value
\begin{equation}
  \label{e:ro1}
 \rho^{[1]}(\vec{r},\vec{r'}) =
A \langle \Psi | \hat{ \rho}^{[1]} (\vec{r},\vec{r'})
| \Psi \rangle
\end{equation}
of the projection operator $\hat{ \rho}^{[1]} (\vec{r},\vec{r'})$,
\begin{eqnarray*}
\hat{ \rho}^{[1]} (\vec{r},\vec{r'}) 
    & = &  |\vec{r} \rangle_A \mbox{}_A\langle\vec{r'}| \\ 
    &\equiv &
    \int | \vec{r}_1 \ldots \vec{r}_{A-1}\vec{r}\rangle
    \de\vec{r}_1 \ldots \de\vec{r}_{A-1}
    \langle \vec{r}_1 \ldots \vec{r}_{A-1} \vec{r'} |   \\
    &=&
    \exp ({-{\rm i} \hat{\vec{p}}_A\cdot\vec{r}})
    |\vec{r}_A=0\rangle\langle\vec{r}_A=0|
    \exp ({{\rm i} \hat{\vec{p}}_A\cdot\vec{r'}}) \\
    &=&
    \exp ({-{\rm i} \hat{\vec{p}}_A\cdot\vec{r}})
    \delta (\hat{\vec{r}}_A)
    \exp ({{\rm i} \hat{\vec{p}}_A\cdot\vec{r'}})
\end{eqnarray*}
in a given unit-normalized state $\Psi$.
Its diagonal elements give the one-body density
distribution $ \rho (\vec{r}) = \rho^{[1]}(\vec{r},\vec{r}) $.
The off-diagonal elements  $\rho^{[1]}(\vec{r},\vec{r'})$
%with $\vec{r} \ne \vec{r'} $
provide a measure of the correlation between
the probabilities to find a particle
in the two positions  $\vec{r} $ and $ \vec{r'} $
while all the other particles are kept fixed.
Such a definition seems to be satisfactory in the case of infinite systems,
or systems bound by an external potential, e.g. the electrons
of an atom. However,
it is apparently  problematic for finite self-bound
systems like nuclei, where
the constituent particles are localized around their CM due to their interaction.
%to change the position of one particle by $\vec{r'}-\vec{r}$
%while the rest remain fixed means that the center of mass (CM)
%of the system has been shifted by $(\vec{r'}-\vec{r})/A$, which is
%a spurious translation.
%In fact, the r.h.s. of eq.~(\ref{e:ro1})
%contains the complete wavefunction $\Psi$ whose
%constructing should be sufficiently accurate
%to separate properly the CM motion.
Therefore, we prefer to deal with the intrinsic particle distributions
that depend
only on intrinsic wavefunctions and Jacobi coordinates.
Only such quantities are of physical
meaning in the case of finite self-bound 
nonrelativistic systems.
In the next subsections this will be demonstrated
for the intrinsic 1DM and 2DM 
and related quantities.

\subsection{The intrinsic one-body density matrix and momentum distribution}

The intrinsic 1DM
in coordinate space may be defined as
\begin{eqnarray}
\label{e:ROINT1}
\rho^{[1]}_{\rm int} (\vec{r},\vec{r'})
&\!\!\equiv\!\! &
A \langle 
\Psi_{\rm int} |
\hat{\rho}_{\rm int}^{[1]} (\ve{r},\ve{r'})
| \Psi_{\rm int} \rangle 
\\ 
&=& 
  A\langle \Psi_{\rm int} |{ \ve{\xi}}_{A-1} = \ve{r} \rangle
\langle {\ve{\xi}}_{A-1} = \ve{r'} | \Psi_{\rm int} \rangle
 \\
  &=&\! A \int \!\!  
\de^3\xi_1\ldots\de^3\xi_{A-2}
\Psi^{\dagger}_{\rm int}(\ve{\xi}_1,\ldots ,
\ve{\xi}_{A-2},\vec{r}) 
\nonumber \\ 
\label{e:ROINT2}
 & &  \times 
\Psi_{\rm int}(\ve{\xi}_1,\ldots ,
\ve{\xi}_{A-2},\vec{r'})  
,
\end{eqnarray}
so that the normalization condition $\int\de^3 r \rho^{[1]}_{\rm int} (\vec{r},\vec{r}) = A$
is satisfied.
We would like to emphasize that
this is not an ``imposed" definition.
It appears naturally when evaluating the dynamical FF
\cite{ShG1974} 
(or its diagonal part, if one uses the
terminology adopted in Chapter XI of
the monograph \cite{GW64}), which is related to the intrinsic OBMD \cite{KoS1985}
\begin{equation}
\label{e:np1}
\eta_{\rm int} (\vec{p}) \equiv %\eta^{[1]}_{\rm int} (\vec{p}) = 
A \langle \Psi_{\rm int} |
\hat{\eta}_{\rm int}(\vec{p}) | \Psi_{\rm int}\rangle 
%\equiv
%A \langle \hat{\eta}^{[1]}_{\rm int}(\vec{p}) \rangle
\end{equation}
with
\begin{eqnarray} 
\hat{\eta}_{\rm int}(\vec{p}) 
&=&  \delta (\vec{p}-\hat{\vec{p}}_A + \hat{\vec{P}}/A) =
\delta (\vec{p}- \hat{\vec{\eta}}_{A-1} ) 
\label{e:npo}
\\ &=& 
| {\vec{\eta}}_{A-1} =
\vec{p}\rangle \langle {\vec{\eta}}_{A-1}=\vec{p} |
 .
\end{eqnarray}
The OBMD is the Fourier transform of the 1DM $ \rho^{[1]}_{\rm int} (\vec{r},\vec{r'}) $,
\begin{equation}
\label{e:npf}
\eta_{\rm int} (\vec{p}) = (2\pi )^{-3} \int \de^3r\de^3r' \exp{[{\rm i}\vec{p}\cdot
(\vec{r}-\vec{r'})]} \rho_{\rm int}^{[1]}(\vec{r},\vec{r'})
 .
\end{equation}
See also ref.~\cite{CPS84}. 
At  the same time, the intrinsic one-body density
$\rho_{\rm int} (\vec{r}) $
is
the Fourier transform
of the elastic FF determined by eq.~(\ref{e:ff1}), or inversely,
\begin{equation}
\label{e:fff}
F_{\rm int}(\ve{q}) = \frac{1}{A}\int
\e^{{\rm i} \ve{q}\cdot\ve{r} }
\rho_{\rm int} (\ve{r}) \de^3r
 .
\end{equation}
From eq.~(\ref{e:fff}) it follows
that $\rho_{\rm int} (\vec{r}) = A \langle \Psi_{\rm int}| \hat{\rho}_{\rm int} (\vec{r})
| \Psi_{\rm int} \rangle $, where
\begin{equation}
\label{e:rhor}
\hat{\rho}_{\rm int} (\ve{r}) =  \delta ( \ve{r} - \hat{\ve{r}}_A + \hat{\ve{R}} )
=  \delta ( \ve{r} - \frc{A-1}{A}\hat{\ve{\xi}}_{A-1} )
.
\end{equation}
We notice that
\begin{equation}
\label{e:rhorel}
\rho_{\rm int} (\vec{r}) =
\left[\frc{A}{A-1}\right]^3
\rho^{[1]}_{\rm int} (\frc{A}{A-1}\vec{r},\frc{A}{A-1}\vec{r})
 .
\end{equation}
In other words, the intrinsic 1DM does not have the property
$ \rho^{[1]}(\vec{r}) = \rho^{[1]}(\vec{r},\vec{r})$ which can be
justified for infinite systems, although it has often
been exploited in approximate
treatments of finite systems
(cf., however, ref.~\cite{VNW1998},
where an alternative
definition of the 1DM for finite self-bound
systems was proposed).

\subsection{The intrinsic two-body density matrix
and two-body momentum distribution}

The formulation presented above will now be
applied to the 2DM. In particular,
we will focus on the TBMD,
usually defined as the
diagonal part of the 2DM in momentum space \cite{PMK2003,Pap2004}.
As we have already discussed,
the relevant definitions require some revision in the
case of finite, self-bound systems.
Here we will consider
the expectation value
\begin{eqnarray} 
{\eta}^{[2]}_{\rm int}(\ve{p},\ve{k}) 
&=& A(A-1)
\langle \Psi_{\rm int} | \delta ( {\hat{\ve{p}}}_{A-1} -
\frc{1}{A}\hat{\ve{P}} - \ve{p} )
\nonumber \\ && \times \delta ( {\hat{\ve{p}}}_A -
\frc{1}{A}\hat{\ve{P}}  - \ve{k} )
| \Psi_{\rm int} \rangle 
\nonumber \\  
\label{e:n21}
 &\equiv& 
A(A-1) 
\langle \Psi_{\rm int} | 
{{\hat{\eta}}^{[2]}_{\rm int}}(\ve{p},\ve{k}) 
| \Psi_{\rm int} \rangle
,
\end{eqnarray}
that can be interpreted as the TBMD
with respect to the intrinsic momentum variables.
%Just such a distribution occurs when
%evaluating the differential
%cross sections of the exclusive $(e,e'NN)$
%reaction on nuclei within the PWIA
%and under
%some additional simplifications.
We can write for the operator ${\hat{\eta}^{[2]}_{\rm int}}(\ve{p},\ve{k})$
\begin{equation}
\label{e:n22}
%{\hat{\eta}_{\rm int}}(\ve{p},\ve{k}) \equiv 
{{\hat{\eta}}^{[2]}_{\rm int}}(\ve{p},\ve{k})
 = (2\pi )^{-6} 
\int \de^3\lambda_1 \de^3\lambda_2
\e^{-{\rm i}\vec{p}\cdot\vec{\lambda}_1}
\e^{-{\rm i}\vec{k}\cdot\vec{\lambda}_2}
\hat{E}_{\rm int} (\vec{\lambda}_1 , \vec{\lambda}_2)
.
\end{equation}
The operator
$\hat{E}_{\rm int} (\vec{\lambda}_1 , \vec{\lambda}_2)$
is expressed in terms of the Jacobi variables,
\begin{equation}
\label{e:Eofn2}
\hat{E}_{\rm int} (\vec{\lambda}_1 , \vec{\lambda}_2) =
\exp [{\rm i} \vec{\lambda}_1 \cdot{\hat{\ve{\eta}}}_{A-2} ]
\exp [{\rm i} ( \vec{\lambda}_2 - \mbox{$\frac{1}{A-1}$} {\vec{\lambda}_1} )
\cdot{\hat{\ve{\eta}}}_{A-1} ]    ,
\end{equation}
if the relations
$
{\hat{\ve{p}}}_A -\hat{\vec{P}}/A= {\hat{\ve{\eta}}}_{A-1}$ and
${\hat{\ve{p}}}_{A-1} - {\hat{\ve{p}}}_A = {\hat{\ve{\eta}}}_{A-2} - \frac{A}{A-1} {\hat{\ve{\eta}}}_{A-1}
 $ are used.
Using the completeness of the $\ve{\xi}$ -- basis, we find
\begin{eqnarray}
\hat{E}_{\rm int} (\vec{\lambda}_1 ,
\vec{\lambda}_2)
&=&
 \int \de^3x\de^3y \de^3x'\de^3y'
\delta (\ve{x} + \vec{\lambda}_1 - \ve{x'} )
 \nonumber \\
& \times &   
\delta (\ve{y} + \vec{\lambda}_2 -
\frc{1}{A-1} \vec{\lambda}_1 - \ve{y'} ) 
\nonumber \\ 
& \times &   
{\hat{\rho}^{[2]}_{\rm int}} (\vec{x},\vec{y};\vec{x'},\vec{y'})
\label{e:EJac}
\end{eqnarray}
with
\begin{eqnarray*}
{\hat{\rho}^{[2]}_{\rm int}} (\vec{x},\vec{y};\vec{x'},\vec{y'})
 &=&
|{\vec{\xi}}_{A-2}=\vec{x}\rangle\langle
{\vec{\xi}}_{A-2}=\vec{x'}|
 \nonumber \\
& & \otimes
|{\vec{\xi}}_{A-1}=\vec{y}\rangle\langle
{\vec{\xi}}_{A-1}=\vec{y'}| \ .
\end{eqnarray*}
The latter is the intrinsic 2DM operator in coordinate space.
It follows from eq.~(\ref{e:EJac}) that
\begin{eqnarray}
{\hat{\eta}^{[2]}_{\rm int}}(\ve{p},\ve{k}) &=&
(2\pi )^{-6} \int \de^3x\de^3y \de^3x'\de^3y'
{\hat{\rho}^{[2]}_{\rm int}}
(\vec{x},\vec{y};\vec{x'},\vec{y'})  .
\nonumber \\ && \times 
\exp [{\rm i} ( \ve{p} +
\frc{1}{A-1}\ve{k} ) \cdot(\ve{x} - \ve{x'}) ] \
\nonumber \\
 && \times
\exp [{\rm i} \ve{k} \cdot(\ve{y} - \ve{y'}) ]
\label{e:etaintop}
\end{eqnarray}
Unlike the usual relationship
\begin{eqnarray*} 
{\hat{\eta}}^{[2]}    (\vec{p},\vec{k}) 
&\equiv&
{\hat{\eta}}^{[2]} (\vec{p},\vec{k};\vec{p},\vec{k}) 
\\ &=& 
(2\pi )^{-6} \int \de^3r\de^3s \de^3r'\de^3s'
\e^{{\rm i} \vec{p}\cdot(\vec{r} - \vec{r'}) }
\e^{{\rm i} \vec{k}\cdot(\vec{s} - \vec{s'}) }
\\ && \times {\hat{\rho}}^{[2]} (\vec{r},\vec{s};\vec{r'},\vec{s'})
 ,
\end{eqnarray*} 
where
${\hat{\eta}}^{[2]}  (\vec{p},\vec{k})$ is the TBMD operator
and
${\hat{\rho}}^{[2]} (\vec{r},\vec{s};\vec{r'},\vec{s'})$
(${\hat{\eta}}^{[2]} (\vec{p},\vec{k};\vec{p'},\vec{k'}) $)
the 2DM operator in coordinate (momentum) space as defined,
for example, in
refs.~\cite{AHP93,PMK2003},
the r.h.s. of eq.~(\ref{e:etaintop}) contains a shift
$\ve{k}/(A-1)$ of the argument $\ve{p}$, which
may be negligibly small when the particle number $A$ increases. However, this is not
the case for few-body systems.

\section{The intrinsic density matrices and related quantities
%Relevant operators
in the Cartesian representation}

%Each of the operators
The intrinsic quantities are defined above in terms of operators, which
can be written as products of $A$ operators acting
on the subspaces of the separate $A$ particles.
Here we will show their evaluation in the Cartesian representation.

As an illustration, let us start from
the operator related to the elastic FF, eq.~(\ref{e:ff1}),
$$
\hat{F}_{\rm int} (\ve{q}) =
\exp [ {\rm i}\ve{q} \cdot (\hat{\ve{r}}_A - \hat{\ve{R}} )] =
\e^{-{\rm i} \frac{\hat{\ve{r}}_1 }{A}
\cdot \ve{q}} \ldots
\e^{-{\rm i} \frac{\hat{\ve{r}}_{A-1} }{A}
\cdot \ve{q}}
\e^{ {\rm i} \frac{A}{A-1} \hat{\ve{r}}_{A}
\cdot \ve{q}}
.
$$
Notice that
$$
{\hat {\rho}}_{\rm int} (\ve{r}) =
\delta ( \hat{\ve{r}}_A - \hat{\ve{R}} - \ve{r} )
= (2\pi)^{-3}\int \e^{- {\rm i} \ve{q}\cdot \ve{r} }
\hat{F}_{\rm int} (\ve{q}) \de^3q 
%\,\,  \hat{F}_{\rm int} (\ve{q}) =
%\exp [ {\rm i}\ve{q} (\hat{\ve{r}}_A -
%\hat{\ve{R}} )]
 .
$$
Using eqs.~(\ref{e:cao}), (\ref{e:com1})
and (\ref{e:vexp}), we find
$$
\e^{-{\rm i} \frac{\hat{\ve{r}} }{A}
\cdot \ve{q}} = \e^{-{\rm i} \frac{r_0}{\sqrt{2} A} \ve{q}
\cdot (\hat{\vec{a}}^{+} + \hat{\vec{a}} )} 
%$$ $$
= \e^{  - \frac{r_0^2 {q}^2}{4 A^2} }
\e^{ -{\rm i} \frac{r_0}{\sqrt{2} A}
\ve{q} \cdot \hat{\vec{a}}^{+} }
\e^{  -{\rm i} \frac{r_0}{\sqrt{2} A}
\ve{q} \cdot \hat{\vec{a}} }  \ ,
$$
$$
\e^{ {\rm i} \frac{A-1}{A} \ve{r} \cdot \ve{q} } =
\e^{  - \frac{(A-1)^2}{A^2} \frac{r_0^2 {q}^2}{4 } }
\e^{ {\rm i} \frac{A-1}{A}
\frac{r_0}{\sqrt{2} } \ve{q} \cdot \hat{\vec{a}}^{+} }
\e^{  {\rm i} \frac{A-1}{A}
\frac{r_0}{\sqrt{2} } \ve{q} \cdot \hat{\vec{a}} } \ ,
$$
whence
$$
{\hat{F}}_{\rm int} (\ve{q}) =
\e^{-(1-\frac{1}{A})\frac{r_0^2 {q}^2}{4}}
\hat{O}_1(\vec{\alpha}) \ldots \hat{O}_{A - 1}
(\vec{\alpha}) \hat{O}_A(\vec{\beta}) \ ,
$$
where
\begin{equation}
\label{e:2.5}
\vec{\alpha} = - {\rm i} \frc{r_0}{\sqrt{2}A} \vec{q}
\hspace{7mm},\hspace{7mm}
\vec{\beta} = {\rm i} \frc{r_0}{\sqrt{2} }
( 1 - A^{-1} ) \vec{q} .
\end{equation}
Henceforth we use the notation
\begin{equation}
\label{e:oops}
\hat{O}_{i}(\vec{z})
\equiv
\e^{-\vec{z}^{\ast} \cdot \hat{\vec{a}}^{+}_{i} }
\e^{ \vec{z} \cdot \hat{\vec{a}}_{i} }
 ,
\end{equation}
with $ i = 1,2, \ldots, A $ and any complex vector $\vec{z}$,
for a recurring operator structure.
In fact, one finds that   
all the intrinsic one- and two-particle operators of interest 
contain this operator structure: 
\[
\hat{O}_1(\vec{z})
\cdots
\hat{O}_{A-2}(\vec{z})
\hat{O}_{A-1}(\vec{x}_2)
\hat{O}_A(\vec{x}_1)
 , \]
where the vectors $\vec{z}, \vec{x}_2, \vec{x}_1$ 
are related by the equation  
$ (A-2)\vec{z} + \vec{x}_2 + \vec{x}_1 = 0 $. 
In the case of one-particle operators, $\vec{x}_2=\vec{z}$.  

Therefore, according to eq.~(\ref{e:ff1}), 
the elastic FF can be written as
\begin{equation}
\label{e:2.7}
F_{\rm int}(\vec{q})
= \e^{-(1-\frac{1}{A})\frac{r_0^2 q^2}{4}}
\langle \Psi_{\rm int} |
\hat{O}_1(\vec{\alpha}) \ldots \hat{O}_{A - 1}
(\vec{\alpha}) \hat{O}_A(\vec{\beta})
| \Psi_{\rm int} \rangle \ ,
\end{equation}
where the vectors $\ve{\aco}$ and $\ve{\beta}$ are
determined by
eqs.~(\ref{e:2.5}). 
%for $\ve{q} = \ve{q}$ .
In the r.h.s. of the equation we find
the Tassie-Barker (TB) factor
$\exp (r_0^2 q^2/4A)$
\cite{TB58}
with the ``length" parameter $r_0$.
In our approach this factor results
from the specific structure
of the intrinsic operator
${\hat{F}}_{\rm int} (\ve{q})$,
being independent of the nuclear structure (in general,
the structure of the finite system under study).
As it is well known, the TB factor appears directly in calculations,
where the nuclear ground state is described by
the simple harmonic oscillator model.

In the fixed--CM approximation,
as we demonstrated in Sec.~2 by means of 
eqs.~(\ref{e:estoin1}), (\ref{e:estoin}),
one has to evaluate the ratio
\begin{equation}
\label{e:2.8}
F_{\rm EST} (\vec{q}) = \frac { \langle \Psi_0 |
\hat{F}_{\rm EST}
( \ve{q} ) | \Psi_0 \rangle }
        { \langle \Psi_0 |\hat{F}_{\rm EST} (0)
| \Psi_0 \rangle }
\end{equation}
with
$$
\hat{F}_{\rm EST} ( \ve{q} ) =
%(2\pi)^3
\delta (\hat{\ve{R} } ) {\hat{F}}_{\rm int} (\ve{q}) .
$$
Using the integral representation
$(2\pi)^3 \delta (\hat{\ve{R} } ) =
\int \exp({\rm i} \ve{\xi}\cdot \hat{\ve{R}} ) \de^3 \xi $
and applying the same technique as before,
one can show that
\begin{eqnarray}
\hat{F}_{\rm EST} ( \ve{q} ) 
&=& 
\e^{-(1-\frac{1}{A})\frac{r_0^2 {q}^2}{4}}
 \int \de^3 \xi \e^{ - r_0^2 {\xi}^2/4A }
\nonumber \\ && \times 
\hat{O}_1(\vec{\alpha}') \ldots
\hat{O}_{A - 1}(\vec{\alpha}') \hat{O}_A(\vec{\beta}') \  ,
%\eqno(2.9)
\label{e:2.9}
\end{eqnarray}
where
$$
\vec{\alpha}'= {\rm i}\frc{r_0}{\sqrt{2}A}
(\vec{\xi} - \vec{q} )
\hspace{3mm},\hspace{3mm}
\vec{\beta}' = {\rm i}\frc{r_0}{\sqrt{2} }
[ \ve{\xi} + ( 1 - A^{-1} ) \vec{q} ] \ .
$$

When calculating expectation values
like those in eq.~(\ref{e:2.8}), the representation
(\ref{e:2.9}) is
especially helpful if the wavefunction $\Psi_0 $
is a Slater determinant or a linear combination
of Slater determinants.
This has been demonstrated in particular in
refs.~\cite{DOS1975,KoS1977},
where the single-particle orbitals entering the
Slater determinant are eigenfunctions of a harmonic-oscillator potential
with oscillator parameter $r_0$.
Recently,
similar calculations have been carried out beyond the HOM
\cite{GKS2002} with the single-particle orbitals
approximated by a truncated expansion
in the Cartesian basis vectors of eq.~(\ref{e:bas}).

The intrinsic 1DM operator can also be expressed
in terms of the Cartesian operators
$\hat{\vec{a}}^{+} $ and $ \hat{\vec{a}} $.
We can rewrite
the operator $\hat{\rho}_{\rm int}^{[1]} (\ve{r},\ve{r'}) $
from eq.~(\ref{e:ROINT1}) as
\begin{eqnarray}
{\hat{\rho}}^{[1]}_{\rm int} (\vec{r},\vec{r'})
&=&
\e^{-{\rm i} \hat{\ve{\eta}}_{A-1} \cdot\ve{r} }
\delta (\hat{\ve{\xi}}_{A-1})
\e^{ {\rm i}
\hat{{\ve{\eta}}}_{A-1} \cdot\ve{r'} }
%= (2\pi)^{-3} \int \de^3 \lambda
%\hat{Z}_{\rm int} (\ve{r},\ve{r'}; \ve{\lambda}) \ ,
\nonumber  \\
%\hat{Z}^{\rm int} (\ve{r},\ve{r'}; \ve{\lambda})
%&\equiv &
&=& 
(2\pi)^{-3} \int \de^3 \lambda
\exp [-{\rm i} (\hat{\vec{p}}_A -
\frc{\hat{\vec{P}}}{A} )\cdot\ve{r} ]
\nonumber  
 \\ && \times \exp [{\rm i} \frc{A}{A-1}
\ve{\lambda}\cdot(\hat{\ve{r}}_A - \hat{\ve{R} } ) ]
 \nonumber \\
&&
\times \exp [{\rm i} (\hat{\vec{p}}_A -
\frc{\hat{\vec{P}}}{A} )\cdot\ve{r'} ] \, .
\label{e:2.11}
\end{eqnarray}
As before, the general idea is to bring this operator  
%$\hat{Z}^{\rm int}  (\ve{r},\ve{r'}; \ve{\lambda}) $
in a form with normal ordering, where
the destruction operators $\hat{\vec{a}}_A$
are to the right with respect to the creation
operators $ \hat{\vec{a}}^{+}_A $.
To do this, we note that
\begin{eqnarray*}
\exp \left[{\rm i} \frc{A}{A-1}
\ve{\lambda}\cdot(\hat{\ve{r}}_A - \hat{\ve{R} } ) \right]
&=&
\exp [ - \frc{A}{A-1} \frc{r_0^2 {\lambda}^2}{4} ] 
\\ && \times 
\exp [{\rm i} \frc{A}{A-1} \frc{r_0}{\sqrt{2} A}
\ve{\lambda} \cdot
(\hat{\vec{a}}^{+}_A - \frc{\hat{\ve{D}}^{+}}{A} ) ]
 \\ && \times
\exp [{\rm i} \frc{A}{A-1} \frc{r_0}{\sqrt{2} A}
\ve{\lambda} \cdot
(\hat{\vec{a}}_A - \frc{\hat{\ve{D}}}{A} ) ]  \, ,
\end{eqnarray*}
where  $\hat{\ve{D}} = \sum_{\alpha =1}^A
\hat{\ve{a}}_{\alpha} $
is the ``collective"
destruction operator with the property
$[\hat{D}_l, \hat{D}^{+}_j] = A \delta_{lj}\,\,
(l,j = 1,2,3)$.
After some modest effort we arrive at
the following result:
\begin{eqnarray*}
\hat{\rho}^{[1]}_{\rm int} (\vec{r},\vec{r'})
&=&  {(2\pi)^{-3}}  
{\exp [- \frc{A-1}{A} \frc{p_0^2}{4}
(\ve{r} - \ve{r'})^2 ] } 
\\ && \times 
\int \de^3 \lambda
\exp [- \frc{A-1}{A} \frc{r_0^2}{4} {\lambda}^2 ]
\exp [-{\rm i}\frc{\ve{\lambda}}{2}
\cdot (\ve{r} + \ve{r'}) ]  \
\\
&&
\times \exp [ ( \frc{p_0}{\sqrt{2}} (\ve{r} - \ve{r'}) +
{\rm i} \frc{A}{A-1}
\frc{r_0}{\sqrt{2} A} \ve{\lambda} )\cdot
( \hat{\vec{a}}^{+} - \frc{\hat{\ve{D}}^{+}}{A} ) ] \
\\
&&
\times \exp [ ( - \frc{p_0}{\sqrt{2}} (\ve{r} - \ve{r'}) +
{\rm i} \frc{A}{A-1}
\frc{r_0}{\sqrt{2} A} \ve{\lambda} )\cdot
( \hat{\vec{a}} - \frc{\hat{\ve{D}}}{A} ) ]
 \, . %\eqno(46)
\end{eqnarray*}
The operator
$\hat{\rho}^{[1]}_{\rm EST} (\vec{r},\vec{r'})=
%(2\pi)^3
\delta (\hat{\ve{R}}) \hat{\rho}^{[1]}_{\rm int} (\vec{r},\vec{r'}) $
can be represented in a similar way.
Furthermore, taking into account eq.~(\ref{e:npf}),
the corresponding OBMD operator 
$\hat{\eta}_{\mathrm{int}}(\vec{p})$ 
equals
\begin{eqnarray}
\hat{\eta}_{\rm int}(\ve{p}) &=&
(2\pi)^{-3} \int \de^3 y \exp [i \ve{p}\cdot\ve{y}]
\exp [- \frc{A-1}{A} \frc{p_0^2}{4} y^2  ]
\nonumber \\
 & &
\times  
\exp [ \frc{p_0}{\sqrt{2}} \ve{y} \cdot
( \hat{\vec{a}}^{+} - \frc{\hat{\ve{D}}^{+}}{A} ) ]
\nonumber \\
 & &
\times  
\exp [ \frc{p_0}{\sqrt{2}} \ve{y} \cdot
( \hat{\vec{a}} - \frc{\hat{\ve{D}}}{A} ) ]
 \\  %\eqno(47) .
&=& 
(2\pi)^{-3} \int \de^3 y \exp [i \ve{p}\cdot\ve{y}]
\exp [- \frc{A-1}{A} \frc{p_0^2}{4} y^2  ]
\nonumber \\
 & &
\times  
\hat{O}_1(\vec{\chi}) \ldots
\hat{O}_{A - 1}(\vec{\chi}) \hat{O}_A(\vec{\gamma }) \, , 
\end{eqnarray} 
As for the operator 
$\hat{\eta}_{\mathrm{EST}}(\vec{p})$, we find 
\begin{eqnarray} 
\hat{\eta}_{\rm EST}(\ve{p}) 
&=& 
(2\pi )^{-3} 
\int \de^3y\de^3\xi 
\exp [- \frc{A-1}{A} \frc{p_0^2}{4} y^2  ]
\nonumber \\ 
& & \times   
\exp [- \frc{1}{A} \frc{r_0^2}{4} \xi^2  ] 
\exp [{\rm i}\ve{y}\cdot\ve{p}] 
\nonumber \\ 
& & \times   
\hat{O}_1(\vec{\chi'}) \ldots
\hat{O}_{A - 1}(\vec{\chi'}) \hat{O}_A(\vec{\gamma '}) \, , 
\label{e:etaest} 
\end{eqnarray}
where 
\begin{equation} 
\vec{\gamma } = 
-\frc{p_0}{\sqrt{2}}(1-\frc{1}{A})\vec{y}  
\hspace{3mm},\hspace{3mm}
\vec{\chi}= 
\frc{p_0}{\sqrt{2}A}\vec{y} 
\label{e:etaintos} 
\end{equation} 
\begin{equation} 
\vec{\gamma '} = {\rm i}\frc{r_0}{\sqrt{2}A }\vec{\xi} - 
\frc{p_0}{\sqrt{2}}(1-\frc{1}{A})\vec{y}  
\hspace{3mm},\hspace{3mm}
\vec{\chi'}= {\rm i}\frc{r_0}{\sqrt{2}A}\ve{\xi} + 
\frc{p_0}{\sqrt{2}A}\vec{y} \, .  
\label{e:etaestos} 
\end{equation} 
With the help of eqs.~(\ref{e:estoin1}), (\ref{e:estoin}), 
(\ref{e:etaest}), (\ref{e:etaestos}), 
the OBMD
\begin{eqnarray*} 
\eta_{\rm EST}(\ve{p}) &=& 
%(2\pi )^3
A \frac{\langle \Psi_0 |
\hat{\eta}_{\rm EST}(\ve{p}) 
%\hat{\eta}_{\rm int}(\ve{p}) \delta (\hat{\vec{R}})
| \Psi_0 \rangle
}{
\langle \Psi_0 | \delta (\hat{\vec{R}})
 | \Psi_0 \rangle }  \\ 
\end{eqnarray*} 
can be evaluated 
 like the elastic FF $F_{\rm EST}(q)$ 
in the fixed-CM approximation.

By comparing the expressions 
$$
\hat{\eta}_{\rm int}(\vec{p}) = (2\pi )^{-3} \int d\vec{y} 
\exp [{\rm i}\vec{y}
\cdot\vec{p}] \exp [ {\rm i}\vec{y} \cdot (\hat{\vec{p}}_A - \frac{1}{A} 
\hat{\vec{P}} )]
, $$ 
following from eq.~(\ref{e:npo}), 
and 
$$
\hat{F}_{\rm int} (\vec{y}) =
\exp [ {\rm i}\vec{y} \cdot (\hat{\vec{r}}_A - \hat{\vec{R}} )]
, $$
we realize that it is sufficient to apply our algebraic technique to
the operator 
$$
\exp [ {\rm i}\vec{y} \cdot (b_1 (\hat{\vec{a}}^{+}_A - 
\frac{1}{A} \hat{\vec{D}}^{+} )
+ b_2 (\hat{\vec{a}}_A - \frac{1}{A} \hat{\vec{D}} ) )]
.$$ 
Then we can generate the intrinsic operators 
$\hat{F}_{\rm int}$ (or $\hat{\rho}_{\rm int}$) and
$\hat{\eta}_{\rm int}$ 
by changing the values of $b_1$, $b_2$. 

Similar manipulations lead to the following expression
for the operator $\hat{E}_{\rm int}$ that enters in the definition of the TBMD operator
${\hat{\eta}^{[2]}_{\rm int}}(\ve{p},\ve{k})$
(see eqs.~(\ref{e:n22})-(\ref{e:EJac})):
\begin{eqnarray}
\hat{E}_{\rm int} (\vec{\lambda}_1 , \vec{\lambda}_2) 
&=\!\!&  \e^{ -\frac{p_0^2\lambda^2}{8} }
\e^{ - \frac{A-2}{A} \frac{p_0^2\Lambda^2}{2} }
\nonumber \\ &\!\! \times & \! \! \hat{O}_1(\vec{\zeta}) \ldots \hat{O}_{A-2}
(\vec{\zeta}) \hat{O}_{A-1}
(\vec{\gamma_2}) \hat{O}_A(\vec{\gamma_1}),
\end{eqnarray}
where
$$
\vec{\gamma_1} = \frc{p_0}{\sqrt{2}} (\frc{A-2}{A} \vec{\Lambda} -
\frc{1}{2} {\ve{\lambda}} )  
\,\, , \,\,
\vec{\gamma_2} = \frc{p_0}{\sqrt{2}} (\frc{A-2}{A}
\vec{\Lambda} + \frc{1}{2} {\ve{\lambda}} ) 
\,\, , 
$$ 
$$ 
\vec{\zeta} = - \sqrt{2} \frc{p_0}{A} \vec{\Lambda}
 .
$$
We have set $\vec{\Lambda}=(\vec{\lambda}_1+\vec{\lambda}_2)/2$
and
$\vec{\lambda}=\vec{\lambda}_1-\vec{\lambda}_2$.
In order to obtain the TBMD operator in the fixed--CM approximation,
one first needs to evaluate
\begin{equation}
\hat{E}_{\rm EST}  (\vec{\lambda}_1 , \vec{\lambda}_2) = (2\pi)^3 \delta(\hat{\ve{R}})
\hat{E}_{\rm int} (\vec{\lambda}_1 , \vec{\lambda}_2)
 . %\eqno(49)
\end{equation}
Again, after some algebra one can show that
$$
\hat{E}_{\rm EST}  (\vec{\lambda}_1 , \vec{\lambda}_2) =
\int \de^3\kappa \e^{- r_0^2 \kappa^2/4A} \e^{ - {p_0^2\lambda^2 / 8} }
\e^{ - \frac{A-2}{2A} {p_0^2\Lambda^2} }
$$
\begin{equation}
 \times
\hat{O}_1(\vec{\zeta}') \ldots \hat{O}_{A-2}
(\vec{\zeta}') \hat{O}_{A-1}
(\vec{\gamma}_2') \hat{O}_A(\vec{\gamma}_1')  \ ,
\end{equation}
where
$$
\vec{\gamma}_1' = {\rm i}\frc{ r_0}{\sqrt{2}A}\vec{\kappa} +
\frc{p_0}{\sqrt{2}} (\frc{A-2}{A} \vec{\Lambda} -
\frc{1}{2} {\ve{\lambda}} ) \, , 
$$
$$
\vec{\gamma}_2' = {\rm i}\frc{  r_0}{\sqrt{2}A}\vec{\kappa} +
\frc{p_0}{\sqrt{2}} (\frc{A-2}{A} \vec{\Lambda} +
\frc{1}{2} {\ve{\lambda}} ) 
$$
and
$$
\vec{\zeta}' = {\rm i}\frc{ r_0}{\sqrt{2}A}
\vec{\kappa} - \sqrt{2}
\frc{p_0}{A} \vec{\Lambda} \ .
$$
The respective TBMD can be written as
\begin{eqnarray}
\eta^{[2]}_{\rm EST} (\vec{p},\vec{k}) 
&=&
(2\pi )^{- 6} A(A-1)\int \de^3\Lambda \de^3\lambda
\e^{-{\rm i}\vec{p}\cdot(\vec{\Lambda}+\vec{\lambda}/2) }
\nonumber \\ && \times 
\e^{-{\rm i}\vec{k}\cdot(\vec{\Lambda}-\vec{\lambda}/2) }
\frac{ N(\vec{\Lambda}+\vec{\lambda}/2,
\vec{\Lambda}-\vec{\lambda}/2  ) }
{N(0,0)}
\label{e:49}
\end{eqnarray}
with
$$
 N(\vec{\lambda_1},\vec{\lambda_2} ) = \langle \Psi_0 |
\hat{E}_{\rm EST}  (\vec{\lambda}_1 , \vec{\lambda}_2)
| \Psi_0 \rangle
 . $$
One can verify that this distribution meets the sequential relation
\begin{equation}
\label{e:seq1}
\int \eta^{[2]}_{\rm EST} (\vec{p},\vec{k}) \de^3 p = (A - 1) \eta_{\rm EST} (\vec{k})
.
\end{equation}

The exposed method can be helpful in more general 
situations when one has to handle
translationally invariant states of the kind
\begin{equation}
\label{e:g1}
| \Psi_{\ve{P}}^{\rm G} \rangle = | \ve{P} )\ | 
\Psi_{\rm int}^{\rm G} \rangle ,  
\end{equation}
with  
the normalized state vector in the $(3A - 3)-$dimensional 
intrinsic Hilbert space 
$|\Psi_{\rm int}^{\rm G} \rangle $ defined by 
\begin{equation}
\label{e:gint}
|\Psi_{\rm int}^{\rm G} \rangle =
\frac{ (\rm G |\Psi_0 \rangle }
{ [{\langle \Psi_0 |\rm G )(\rm G |\Psi_0 \rangle }]^{1/2} }
 . \end{equation}
Here, $|\rm G )$ is any arbitrary vector in the CM
space so that the scalar product $(\rm G |\Psi_0 \rangle $  
represents integration of
the CM variable only. 
Such general cases were considered in ref.~\cite{EST1973}. 
The expectation value 
of the intrinsic operator $\hat{A}_{\rm int}$ 
in the state
$
|0_{\rm G} \rangle = \int \de^3P c(\vec{P}) | \ve{P} )\ |\Psi_{\rm int}^{\rm G} \rangle
$ is  
\begin{equation}
%\label{e:gex}
A_{\rm G} \equiv
\langle 0_G | \hat{A}_{\rm int} |0_G \rangle = {\langle \Psi_0|
 \hat{A}_{\rm G}
 | \Psi_0 \rangle } / { \langle\Psi_0 | \rm G )( \rm G |\Psi_0 \rangle }
,
\end{equation}
\begin{equation}
%\label{e:}
\hat{A}_{\rm G} =
| \rm G ) \hat{A}_{\rm int} ( \rm G | = | \rm G )( \rm G | \hat{A}_{\rm int} =
\hat{A}_{\rm int} | \rm G )( \rm G | .
\end{equation}
(Cf. eqs.~(24)-(25) in ref.~\cite{EST1973b}.)

The operator $\hat{A}_{\rm G} $  can be represented in the two equivalent forms:
\begin{equation}
\label{e:AGP}
\hat{A}_{\rm G} = \int\de^3 P \int \de^3 P' G^{\ast}(\vec{P'}) G(\vec{P})
 \hat{M} (\vec{P},\vec{P'} )  \hat{A}_{\rm int}
\end{equation}
and
\begin{equation}
\label{e:AGR}
\hat{A}_{\rm G} = \int\de^3 R \int \de^3 R' \tilde{G}^{\ast}(\vec{R'}) \tilde{G}(\vec{R})
 \hat{M} (\vec{R},\vec{R'} )  \hat{A}_{\rm int}
\end{equation}
with the projection operators $ \hat{M} (\vec{P},\vec{P'} ) = |\vec{P})(\vec{P'}| $ and
$ \hat{M} (\vec{R},\vec{R'} ) = |\vec{R})(\vec{R'}| $. The functions
$G(\vec{P}) = (\vec{P} | \rm G ) $ and $\tilde{G}(\vec{R}) = (\vec{R} | \rm G ) $ represent
the vector $| \rm G )$ in the two bases composed of the
$\vec{P}$ - eigenvectors 
(i.e., $\hat{\vec{P}}|\vec{P}) = \vec{P} |\vec{P}) $) and
$\vec{R}$ - eigenvectors (i.e., $\hat{\vec{R}}|\vec{R}) = \vec{R} |\vec{R}) $), 
respectively.

So far, the function $G$ has been totally arbitrary. If we set 
$\tilde{G}(\vec{R}) = \delta (\vec{R}) $, i.e., $| \rm G ) = | \vec{R} = 0 )$ and
$G(\vec{P}) = (2 \pi )^{-3/2} $, we repeat the EST
prescription with the intrinsic state $|\Psi_{\rm int}^{\rm EST} \rangle $,
eq.~(\ref{e:esti}),  and  the expectation value:
\begin{equation}
\label{e:estoin2}
A_{\rm EST} = \frac{\langle \Psi_0| \delta (\hat{\vec{R}}) \hat{A}_{\rm int}
 | \Psi_0 \rangle } { \langle\Psi_0 | \delta (\hat{\vec{R}}) |\Psi_0 \rangle }
\end{equation}
If, on the other hand, we set  $G(\vec{P}) = \delta (\vec{P}) $, i.e.,
$| \rm G ) = | \vec{R} = 0 )$ and $\tilde{G}(\vec{R}) = (2 \pi )^{-3/2} $, we arrive at the
no-fixed-CM approximation after 
Peierls and Yoccoz \cite{PeY1957} in constructing the intrinsic
state
\begin{equation}
\label{e:py}
|\Psi_{\rm int}^{\rm PY} \rangle = \frac{(\vec{P}=0 |\Psi_0 \rangle }
{ [{\langle \Psi_0 | \delta (\hat{\vec{P}}) |\Psi_0 \rangle }]^{1/2} }
 .
\end{equation}
The expectation value of interest is
\begin{equation}
\label{e:pyoin}
A_{\rm PY} = \frac{\langle \Psi_0| \delta (\hat{\vec{P}} ) \hat{A}_{\rm int}
 | \Psi_0 \rangle } { \langle\Psi_0 | \delta (\hat{\vec{P}}) |\Psi_0 \rangle }
 . 
\end{equation}
Note that within the PY approach the TI state vector for a nucleus moving with momentum
$\vec{P}$
is approximated by $|\Psi_{\rm \vec{P}}^{\rm PY} \rangle =
|\vec{P})(\vec{P} |\Psi_0 \rangle $. 
Such a model is incompatible with the Gallileo invariance requirement. 
The 
restoration of Gallileo invariance 
was discussed by Peierls and Thouless \cite{PeT1962}. 
The unit-normalized vector given by
eq.~(\ref{e:py}) corresponds to the 
PY prescription for the nucleus rest frame only.
In this context we would like to 
refer again to the several calculations in nuclei with the ansatz
(\ref{e:py}) performed recently by 
Schmid and his colleagues \cite{Schmid,RGS2004}. %Sch2001,Sch2002a,Sch2002b}. 

Previous experience prompts us 
to unify forthcoming calculations of 
quantities like
(\ref{e:estoin2}) and (\ref{e:pyoin}) 
by introducing one $\delta-$function, 
$\delta ( c_1 \hat{\vec{D}}^{+} + c_2 \hat{\vec{D}}) $,  
where $c_1$, $c_2$ c-number parameters, 
instead of the two functions 
$\delta (\hat{\vec{R}})$  and  $\delta (\hat{\vec{P}})$. 
Along these guidelines, one is able to build up
a family of generating functions which can be calculated with the 
help of 
the algebraic technique developed in this article.

Finally, the following representation
$$
\hat{M} (\vec{R},\vec{R'} ) =\exp ({-{\rm i} \hat{\vec{P}}\cdot\vec{R}})
    \delta (\hat{\vec{R}})
    \exp ({{\rm i} \hat{\vec{P}}\cdot\vec{R'}})
$$
%(cf. the transition below eq.~(\ref{e:ro1}))  
might be useful in calculations 
employing a 
refined determination of the weight function $G$. In particular, 
as outlined in ref.~\cite{EST1973}, 
the ``best" $G$ must be chosen so as to minimize the
expectation value of the intrinsic Hamiltonian. 
See also ref.~\cite{Vin1973}, 
devoted to an optimal separation of CM motion in
many-body systems.

Closing this section, we should note that 
within the simple HOM, where the single-particle states are 
described as pure 
harmonic-oscillator wavefunctions, the total wavefunction is always a product 
of the CM wavefunction and an intrinsic one. 
Therefore, for intrinsic quantities, 
one gets the same results with or without 
the use of a projection technique. 

\section{The intrinsic one-body and two-body momentum distributions: 
Application
to \nuc{4}{He} and discussion}

The many-particle operators encountered so far have much in common with each other
owing to the operator structure appearing in each of them,
\[
\hat{O}_1(\vec{z})
\cdots
\hat{O}_{A-2}(\vec{z})
\hat{O}_{A-1}(\vec{x}_2)
\hat{O}_A(\vec{x}_1)
 . \]
Let us derive their expectation value in the 
independent-particle model. The wavefunction is then a
Slater determinant $|{\rm Det}\rangle$ (eq.~(\ref{Eslat})).
The following formal result is straightforward:
\[ \langle {\rm Det} |
\hat{O}_1(\vec{z})
\cdots
\hat{O}_{A-2}(\vec{z})
\hat{O}_{A-1}(\vec{x}_2)
\hat{O}_A(\vec{x}_1)
|{\rm  Det} \rangle \]
\begin{equation}
\label{e:53b}
=\langle {\rm Det}(-\vec{x}_1,-\vec{x}_2 ,-\vec{z} ) |
 {\rm Det}(\vec{x}_1 ,\vec{x}_2 ,\vec{z} ) \rangle
 , %\eqno(52)
\end{equation}
where
\begin{eqnarray}
\label{e:53}
&& | {\rm Det}(\vec{x}_1  ,\vec{x}_2 ,\vec{z} ) \rangle
\equiv
 \frac{1}{\sqrt{A!}}
\sum_{\hat{\mathcal{P}}\in S_A}
\epsilon_{\mathcal{P}}
\hat{\mathcal{P}}
\{
\e^{\vec{z}\cdot \hat{\vec{a}}_1}
|\phi_{p_1}(1)\rangle
\cdots
\e^{\vec{z}\cdot \hat{\vec{a}}_{A-2}}
\nonumber
\\
& \times &
|\phi_{p_{A-2}}(A-2)\rangle
\e^{\vec{x}_2\cdot \hat{\vec{a}}_{A-1}}
|\phi_{p_{A-1}}(A-1)\rangle
\e^{\vec{x}_1\cdot \hat{\vec{a}}_A}
|\phi_{p_A}(A)\rangle
\} . 
\end{eqnarray}
The r.h.s. of eq.~(\ref{e:53b}) can be
represented as the sum of terms
containing the matrix elements
$\langle \phi_{p_i} | 
%\exp(-\vec{z}^{\ast}
%\cdot \hat{\vec{a}}^{+})
%\exp (\vec{z} \cdot \hat{\vec{a}})
\hat{O}_{m} 
|\phi_{p_j} \rangle $ ($i,j,m = 1, 2, \ldots A  $).
Note that the permutations $\hat{\mathcal{P}}$
are related to the single-particle states
(not the nucleon
labels).
In the simplest case of the $0s^4$ configuration, that we encounter in \nuc{4}{He} nucleus,
all we need is to evaluate the matrix element
$\langle 0s | 
%\exp(-\vec{z}^{\ast} \cdot \hat{\vec{a}}^{+})
%\exp (\vec{z} \cdot \hat{\vec{a}}) 
\hat{O}_{m} 
|0s \rangle $.
We have written
$| \phi_{0s} \rangle = | 0s \rangle
| \sigma \tau \rangle $, where $ | 0s \rangle $ and
$ | \sigma \tau \rangle $  are the space and
spin-isospin parts, respectively,
of $|\phi_{0s}\rangle $.
In the HOM,
$\phi_{0s}(\vec{r})\equiv \langle\vec{r}|0s\rangle
= (\sqrt{\pi}b)^{-{3/2}}\e^{-r^2/2b^2}$,
where $b=r_0=1/p_0$ is the harmonic oscillator parameter.
The state $|0s\rangle$ coincides with the lowest-energy
state $ |000 \rangle $ which is the
``vacuum" of the Cartesian representation, viz.,
$ \hat{\vec{a}} | 000\rangle = 0$.
As a consequence, the radial matrix element of
interest is equal to unity
and
\begin{equation}
\label{e:prod}
\langle 0s^4 |
\hat{O}_1(\vec{z})
\hat{O}_{2}(\vec{z})
\hat{O}_{3}(\vec{x}_2)
\hat{O}_4(\vec{x}_1)
| 0s^4 \rangle = 1,
\end{equation}
rendering the rest of the calculation trivial.
We are now in a position to write down expressions
for all the quantities considered, in the case of
\nuc{4}{He}. In particular, performing the
necessary integrations
and taking into account the normalization of
each quantity, we get for the intrinsic OBMD and TBMD 
in the EST prescription
\begin{eqnarray}
\label{e:54}
\eta_{\rm EST}  (\vec{p}) &=& 
\eta_{\rm EST} (p) = 
\frac{4^{3/2}4b^3}{3^{3/2}\pi^{3/2}}
\e^{-\frac{4}{3}p^2b^2}
 ,
\\
\label{e:55}
\eta^{[2]}_{\rm EST} (\vec{p},\vec{k}) &=&
\frac{2^{3/2}12b^6}{\pi^3}
\e^{-\frac{3}{2}p^2b^2}
\e^{-\frac{3}{2}k^2b^2}
\e^{-\vec{p}\cdot\vec{k}b^2}
 .
\end{eqnarray}
Regarding the properties of these distributions, we find that they satisfy the relation
(\ref{e:seq1}):
$$
\int \eta^{[2]}_{\rm EST} (\vec{p},\vec{k}) \de^3 p = 3 \eta_{\rm EST} (\vec{k})
.
$$
In the simple HOM (with the CM correction not included)
we have \cite{KoV1992,PMK2003,Pap2004}
\begin{eqnarray}
\label{e:56}
\eta (\vec{p}) &=&
\eta (p) =
\frac{4b^3}{\pi^{3/2}}
\e^{-p^2b^2}
 , \\
\label{e:57}
\eta^{[2]}(\vec{p},\vec{k}) &=&
\frac{12b^6}{\pi^3}
\e^{-p^2b^2}
\e^{-k^2b^2}
 .
\end{eqnarray}

We notice that, 
within the HOM, the expression
for the OBMD in the EST approach,
eq.~(\ref{e:54}),
is the same as the one without the CM fixation,
eq.~(\ref{e:56}),
if one substitutes $b$ with $\sqrt{4/3} b$.
This results in a narrowing of
%narrower
distribution $\eta_{\rm EST}  (\vec{p})/ \eta_{\rm EST}  (0)$ compared to
 $\eta (\vec{p}) / \eta(0) $.
But this observation does not mean that the CM correction 
reduces to the
renormalization
$b \to b_{CM} = \sqrt{4/3} b$. In this connection, let us recall the work
\cite{DOS1975},
where the charge FF and the dynamic FF of \nuc{4}{He}
were calculated
under some simplifying 
assumptions using the same oscillatory wavefunction and the EST prescription. Its
main discovery was the 
narrowing of both the intrinsic density and OBMD 
due to the CM correction.
In order to demonstrate this phenomenon, 
let us
write the nucleon density in the EST approach,
\begin{equation}
\label{e:rhoest}
{\rho }_{\rm EST} (r) = 4 \frac{\exp (- \frac{r^2}{{{\bar r}_0}^2})}
{\pi^{3/2} {{\bar r}_0}^3}
,
\end{equation}
that follows directly  
from previous derivations 
and rewrite eq.~(\ref{e:56}) in the form :
\begin{equation}
\label{e:mdest}
{\eta }_{\rm EST} (p) = 4 \frac{\exp (- \frac{p^2}{{{\bar p}_0}^2})}
{\pi^{3/2} {{\bar p}_0}^3}
.
\end{equation}
Here we have come back to the oscillator parameters $r_0$ and $p_0$
and introduced 
the new values ${\bar r}_0 = \sqrt{\frac{A - 1}{A} } r_0$ and
$ {\bar p}_0 = \sqrt{\frac{A - 1}{A} } p_0 $. Meanwhile, we write 
$\sqrt{\frac{A - 1}{A} } $  instead of $\sqrt{3/4}$ to point out a
trend in
the $A$ - dependence of this effect of center-of-mass motion.
We notice that 
the corresponding quantities in the simple HOM
$$
\rho (r) = 4\frac{ \exp (- \frac{r^2}{{r_0}^2} ) } {\pi^{3/2}  {r_0}^3}\ ,
$$
$$
\eta (p) = 4\frac{ \exp (- \frac{p^2}{{p_0}^2} ) } {\pi^{3/2} {p_0}^3}\
$$
are obtained from eqs.~(\ref{e:rhoest}) - (\ref{e:mdest}) when $A \to
\infty $ .

Thus we see (cf. \cite{DOS1975})  that the inclusion of CM corrections
gives rise to
the two renormalizations,
$r_0 \to {\bar r}_0 = \sqrt{\frac{A - 1}{A} } r_0$ and
$p_0 \to {\bar p}_0 = \sqrt{\frac{A - 1}{A} } p_0 $, of the oscillator
parameter
values, $r_0$ and $p_0$, in the density and momentum distributions calculated
within the simple HOM. Evidently, such changes are 
not accounted for by a hasty
replacement of $b$ by $ \sqrt{4/3} b $. 
(Therefore, the discussion in ref.~\cite{CdA1980}, p.~263, 
on the results of ref.~\cite{DOS1975} is incomplete.) 
One may say that we encounter a specific effect of shrinking 
of the density
distribution
$ \rho (r) $ and the momentum distribution $ \eta (p) $. The term
``shrinking" implies that
each of these
densities, after being CM corrected, increases in its central but decreases in
its peripheral
region. 
%(Moreover, the use of the term ``shrinking" instead of ``narrowing" is more
%pertinent since
%$ {\rho }_{\rm EST} (0) \ge \rho (0) $ and $\eta_{\rm 2EST} (0) \ge \eta
%(0)$.)
As it has been shown in the past \cite{DOS1975,DE88}, such a
simultaneous change of the one-body distributions plays an essential role
in getting
a fair treatment of the data on elastic and inelastic electron scattering
off \nuc{4}{He}.
Notice that the product ${\bar r}_0 {\bar p}_0 = 1 - 1/A \ne 1 $, 
unlike the relation $ r_0 p_0 = 1 $. 
The 
commutation rules for the intrinsic coordinates 
$\vec{r'}=\vec{r} - \vec{R} $ 
and conjugate momenta 
$\vec{p'}=\vec{p} - \vec{P}/A $ 
\[ 
[ (\hat{\vec{r'}})_i, (\hat{\vec{p'}})_j ] = \mathrm{i}\delta_{ij}(1-1/A) 
\,\,\, ; \,\,\, i,j = 1,2,3  
\] 
and the corresponding uncertainty principle  
are related to this deviation from unity \cite{KoS1977}.  
Thus, the uncertainty principle is not contradicted 
by the simultaneous shrinking 
of the density and momentum distribution 
(see also \cite{She00}, Lect.~I, Suppl.~C). 

A similar shrinking is encountered 
also in the two-dimensional surface
given by 
the function $\eta^{[2]}_{\gamma {\rm  EST} }(p,k) \equiv 
\eta^{[2]}_{\rm EST}(\vec{p},\vec{k})$  
vs the one given by $\eta^{[2]}_{\gamma} (p,k) \equiv  
\eta^{[2]}(\vec{p},\vec{k}) $,  
at each value of the
correlation angle $\gamma$.
This effect is clearly visible if the TBMD of \nuc{4}{He} 
is represented as
$$
\eta^{[2]}_{\gamma {\rm  EST}}({p},{k}) =\left[ \frc{A}{A-2} \right]^{3/2}
\eta^{[2]}_{\gamma} (p,k)
\e^{ - \frac{1}{A-2} \frac{ (\vec {p} + \vec {k} )^2
}{{p_0}^2} }.
$$

%%%beginABOUTb-new%%% 
Now, in order to compare numerically calculations based upon the formulae
(\ref{e:55}) and
(\ref{e:57}), we adopt, as in ref. \cite{DOS1975}, the following way of
determining the harmonic-oscillator 
parameter $b$. We start with a general expression for the charge FF
$F_C (q) = f_p (q) F (q) $ of a nucleus, where $f_p (q)$ the
correction factor for
the finite proton size.  
The factor $f_p(q)$ is approximated by a Gaussian, 
$f_p (q) = \exp (- {b_p}^2 q^2 )$ 
and for the parameter $b_p$ we choose the value $b_p=0.8$~fm
\cite{KoV1992}.
The FF of the \nuc{4}{He} nucleus equals $F(q) = \exp (-{r_0 }^2 q^2
)$ in the simple
HOM and $F_{\rm EST}(q) = \exp (-{{\bar r}_0 }^2 q^2 )$ in the HOM with CM
corrections.
%Of course, in accordance with eq.~(\ref{e:fff} )  the latter is the Fourier
%transform of the density  ${\rho}_{\rm EST} (r)$ normalized to unity.
In each of the two models the value of the adjustable parameter $r_0 = b$ 
will be determined by the requirement to reproduce the experimental value
of the charge
mean square radius of \nuc{4}{He}, $r_{\rm rms}=1.67$~fm \cite{VJV87}. 
Then in the simple HOM ($b=b_0$) we have
\[ r_{\rm rms}^2=\frac{3}{2} b_0^2 + b_p^2\]
and therefore $b_0=1.197$~fm.
When CM motion corrections are taken into account ($b=b_{\rm CM}$),
\[ r_{\rm rms}^2
= \frac{3}{2} \frac{A-1}{A} b_{\rm CM}^2+b_p^2\]
and therefore $b_{\rm CM}=\frac{A}{A-1}b_0=1.382$~fm.
By adopting this fitting we therefore get the identical $q$-dependence,
$F_{\rm EST}(\vec{q}) = F(\vec{q}) = \exp (-{b_0 }^2 q^2 ) $, 
while the difference
between the respective
OBMDs becomes more considerable.  In fact, we have
$$
{\eta }_{\rm EST} (p) = 4 \left[ \frc{A}{A-1} \right]^3
\frac{{b_0}^3}{\pi^{3/2}}
\exp (- \left[ \frc{A}{A-1} \right]^2 p^2 {b_0}^2 )
$$
vs
$$
\eta (p) = 4  \frac{{b_0}^3}{\pi^{3/2}}
\exp (- p^2 {b_0}^2 ).
$$
The former differs from the latter by a substitution $b_0 \to
\frac{A}{A-1} b_0 $
%%%endABOUTb-new%%% 

In Fig. 1 the TBMD of \nuc{4}{He} is shown for  $p=0, 1, 1.5$~fm$^{-1}$
as a function of $k_p$,
where $\vec{k}$ is parallel to $\vec{p}$
and $\vec{k}=k_p\hat{p}$, i.e., $k_p$
is positive (negative) for
$\vec{k}$ in the same (opposite) direction as $\vec{p}$.
The TBMD that we have calculated using the EST method, i.e.,
with CM motion corrections ($b=b_{\rm CM}$),
is plotted with full lines,
while
the TBMD within the simple HOM with $b=b_0$ is plotted with
long-dashed lines
(``HO1") and
the TBMD within the simple HOM with $b=b_{\rm CM}$
is plotted with short-dashed lines (``HO2") 
taken from refs.~\cite{PMK2003,Pap2004}.
One can observe the shift of the
peak from $k_p=0$ towards negative $k_p$'s,
for $p\neq 0$, due to the
correlation induced by the fixed center of mass.
\begin{figure}
\epsfig{figure=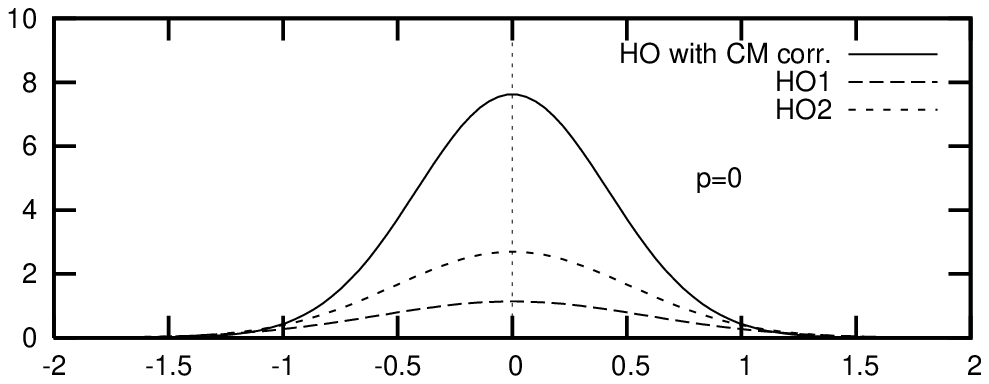,width=80mm}
\epsfig{figure=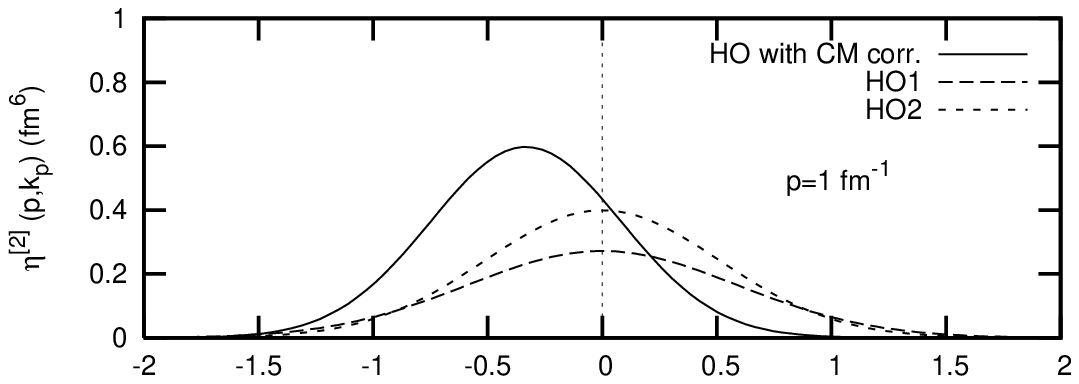,width=80mm}
\epsfig{figure=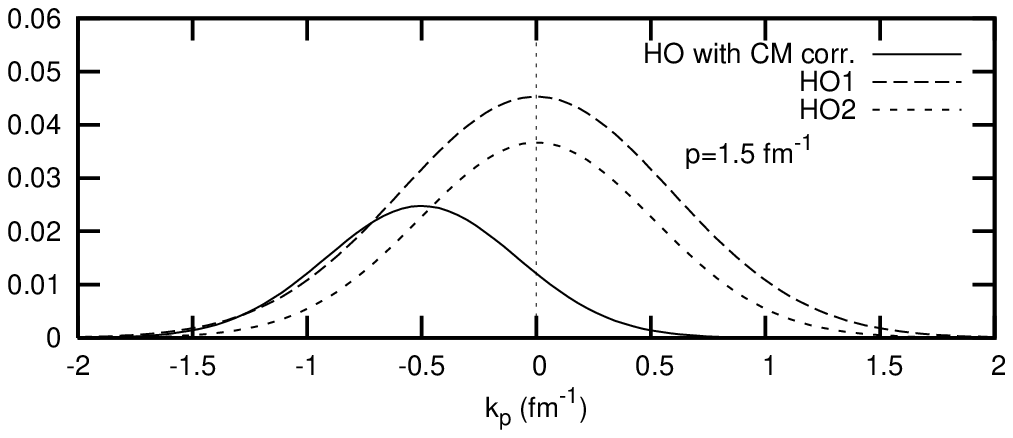,width=80mm}
\caption{%
The TBMD of \nuc{4}{He} for $\vec{p}\parallel \vec{k}$
and $p=$~0,~1,~1.5~fm$^{-1}$,
as a function of $k_p$, where $\vec{k}={k_p}\hat{p}$,
in the HOM model.
Full line: with CM motion effects taken into account
within the EST approach,
eq.~(\ref{e:55}), with $b=b_{\rm CM}$;
long-dashed line:
in the simple HOM, eq.~(\ref{e:57}), with $b=b_0$ (HO1);
short-dashed line:
in the simple HOM, eq.~(\ref{e:57}), with $b=b_{\rm CM}$ (HO2), 
from refs.~\cite{PMK2003,Pap2004}.
} 
\label{f:n2ofk}
\end{figure}

%%%beginABOUTXI%%%
In refs.~\cite{PMK2003,Pap2004}
the dimensionless
quantity
\begin{equation}
\xi (\vec{p},\vec{k})
\equiv
\eta^{[2]}(\vec{p},\vec{k})/\eta (\vec{p})\eta (\vec{k})
\end{equation}
was introduced, as a measure
of correlations of statistical
and dynamical origin
as well as of finite-size effects.
In the complete absence of correlations, $\xi $ should
be equal to $1-1/A$.
In the case of the infinitely extended ideal Fermi gas $\xi $ is defined for $p, k \le
p_F$, where $p_F$ is the Fermi momentum,
%for example,
and if $\vec{p}\neq\vec{k}$, $\xi =1$
(note that $A\rightarrow \infty $), while for
$\vec{p}=\vec{k}$, $\xi=1 -1/\nu$ ($\nu $ is the
degeneracy of the particle states).
Even for the finite non-interacting fermion system,
$\xi =1-1/\nu$ if $\vec{p}=\vec{k}$,
(because
$\eta^{[2]} (\vec{p},\vec{k}) = \eta(\vec{p})\eta(\vec{k})
-\frac{1}{\nu}|\eta_1(\vec{p},\vec{k})|^2$ holds).
Deviations of $\xi (\vec{p},\vec{p}) $
from this value show the effect
of other-than-statistical correlations.
For $\vec{p}\neq\vec{k}$,
deviations from $1-1/A$ is a measure of statistical
and (or) dynamical correlations
in a system of finite size.
In addition, in the case of self-bound systems,
deviations from this value account for the
correlation due to the fixed
center of mass of the system.

\begin{figure}
\epsfig{figure=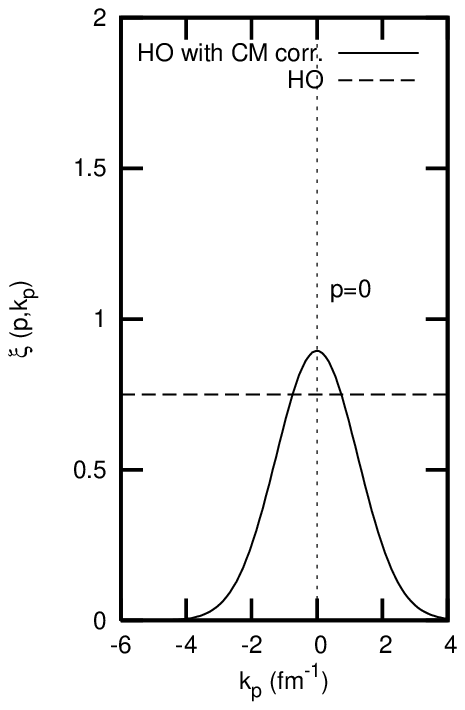,width=28mm}
\epsfig{figure=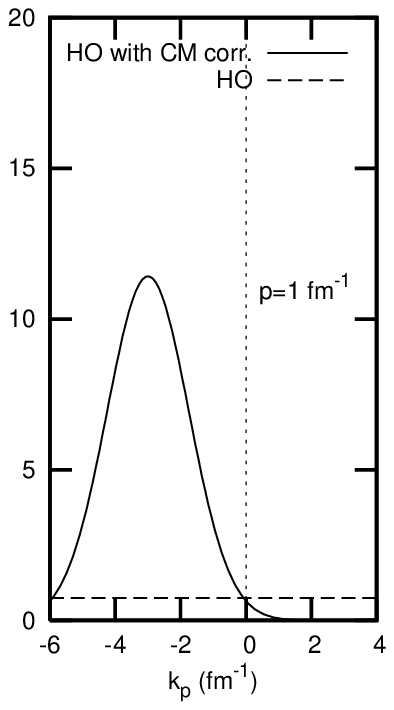,width=28mm}
\epsfig{figure=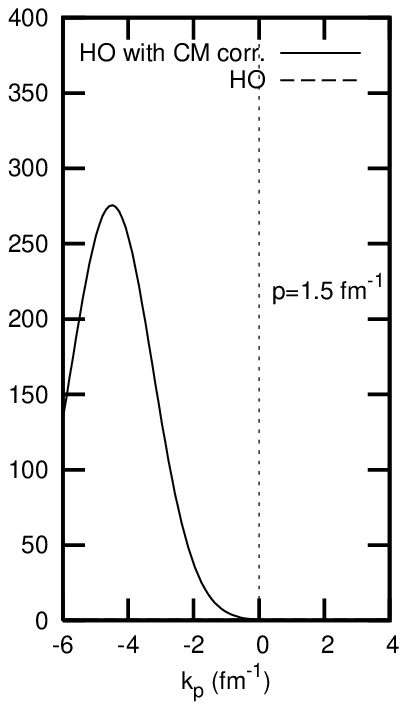,width=28mm}
\caption{%
The quantity $\xi$ of \nuc{4}{He} for $\vec{p}\parallel \vec{k}$
and $p=$~0,~1,~1.5~fm$^{-1}$,
as a function of $k_p$, where $\vec{k}={k_p}\hat{p}$,
in the HOM model.
Full line: with CM motion effects taken into account
within the EST approach, eq.~(\ref{e:60}) for $b=b_{CM}$;
dashed line: in the simple HOM, eq.~(\ref{e:59}), 
refs.~\cite{PMK2003,Pap2004}. 
} 
\label{f:xiof}
\end{figure}

\begin{figure}
\epsfig{figure=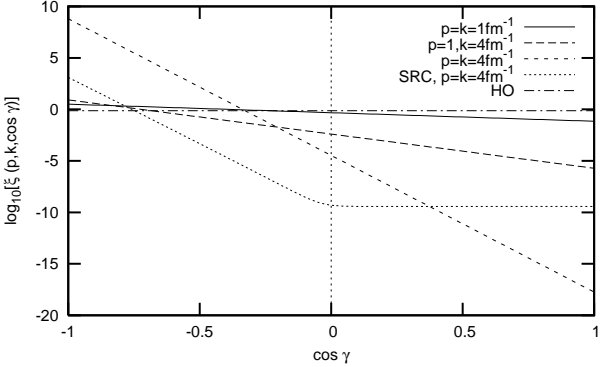,width=80mm}
\caption{%
The quantity $\log_{10}\xi$ of \nuc{4}{He} for
the indicated values of $p$ and $k$
as a function of $\cos{\gamma}$, where
$\gamma$ is the angle between $\vec{p}$ and $\vec{k}$,
in the HOM model.
Full, long-dashed, short-dashed lines:
with CM motion effects taken into account
within the EST approach, eq.~(\ref{e:60}) for $b=b_{CM}$;
dotted line: including short-range correlations within the LOA \cite{PMK2003}, 
without CM motion effects taken into account; 
dot-dashed line: in the simple HOM, eq.~(\ref{e:59})
(result independent of $p$, $k$ and $\gamma$). }
\label{f:xiof2}
\end{figure}

The system of \nuc{4}{He} in the simple HOM is a special case for which
$\nu =A$ and therefore
\begin{equation}
\label{e:59}
\xi =1-1/\nu =0.75
\end{equation}
for all $\vec{p}$ and $\vec{k}$. The same does not
hold after fixing the center of mass. Then we have:
\[ 
\xi_{\rm EST}  (\vec{p},\vec{k})
= \eta^{[2]}_{\rm EST}(\vec{p},\vec{k}) /
\eta_{\rm EST}(\vec{p})
\eta_{\rm EST}(\vec{k})
\] 
\begin{equation}
\label{e:60}
= 0.89493
\e^{-\frac{1}{6}p^2b^2}
\e^{-\frac{1}{6}k^2b^2}
\e^{-\vec{p}\cdot\vec{k}b^2}
\end{equation}
%%%endABOUTXI%%% 
In Fig.~\ref{f:xiof}, we plot $\xi$
as a function of $k_p$, where $\vec{k}=k_p\hat{p}$,
for $p=0, 1, 1.5$~fm$^{-1}$.
In Fig.~\ref{f:xiof2},
$\log_{10}\xi$ is plotted
as a function of $\cos{\gamma}$, where $\gamma$ is the angle
between $\vec{p}$ and $\vec{k}$.
The full, long-dashed and short-dashed lines correspond to the
$\xi_{\rm EST}  (\vec{p},\vec{k})$
of eq.~(\ref{e:60})
for
$p=k=1$~fm$^{-1}$,
for
$p=1$, $k=4$~fm$^{-1}$
and for
$p=k=4$~fm$^{-1}$,
respectively,
and the dashed lines to
$\xi  (\vec{p},\vec{k})=0.75$, eq.~(\ref{e:59}).
The effect of CM correlations is important. The EST TBMD
favors momenta of opposite directions,
as compared to the product of the two OBDM,
while in forward angles $\xi_{\rm EST}$
is significantly reduced.
In refs.~\cite{PMK2003,Pap2004} the effect of short-range correlations (SRC) 
was investigated by including
in the simple HOM Jastrow-type correlations in the calculation of TBMD of the \nuc{4}{He}
nucleus using the lowest-order approximation of 
ref.~\cite{DSB1982}. Significant deviations from
the independent-particle 
picture were found for large values of $p$ and $k$ close to $\gamma = 180^{\circ}$
and $0^{\circ} $ . In Fig.~3 the corresponding quantity $\xi $ for the case
$p=k=4$~{fm}$^{-1} $ is plotted in logarithmic scale for comparison. It is anticipated
that within the EST approach additional corrections due to SRC will appear mainly at high
values of $p$ and/or $k$ and that they will be larger when $\vec{p}$ and $\vec{k}$ are
antiparallel.


\section{Summary and Conclusions }

The intrinsic one-body and two-body density matrices in coordinate space and 
corresponding Fourier transforms in momentum space have been studied for a nucleus 
(a nonrelativistic system) that consists of $A$ nucleons (particles). We have seen how 
these quantities of primary concern can be expressed through expectation values of 
the $A-$particle multiplicative operators 
$\rho_{\rm int}^{[1]}$ and $\rho_{\rm int}^{[2]}$ 
sandwiched between 
intrinsic nuclear states. Our consideration is translationally invariant 
since the operators depend on the relative coordinates and momenta (Jacobi 
variables). To avoid a cumbersome multiple integration, we have developed an 
algebraic technique based upon the Cartesian representation, in which the Jacobi 
variables are the linear combinations of the creation and destruction operators 
% and 
for oscillator quanta in the three different space directions.  
In the framework of the 
subsequent operations the normal ordering of the operators involved in  
$\rho_{\rm int}^{[1]}$ and $\rho_{\rm int}^{[2]}$ 
plays a central role in getting both the general results and the working 
formulae.

The Cartesian representation is convenient and allows us to find simple links between 
the relevant distributions via the generating functions constructed here. In particular, 
the OBMD $\eta(\vec{p})$  and the elastic FF $F(q)$ can be deduced from one and the same 
generating function by changing the values of its arguments.

In the course of such a procedure the so-called Tassie-Barker factors stem directly 
from the intrinsic operators (not the WFs). 
One should emphasize that these factors (different for different
distributions) occur here reflecting  the translationally invariant structure of the
corresponding intrinsic operators. Each of them is a Gaussian whose behavior
in the space of variables is governed by the size parameter $r_0$ (or its reciprocal $p_0$)
and the particle number $A$ for a given system, 
but it does not depend upon the choice of the
g.s. WF. The latter can be a simple Slater determinant, include SRC or not, 
be CM-corrected or not, etc. In practical calculations, such WF's (in particular,
the reference WF $|\Psi\rangle$ in eq.~(\ref{e:esti})) are often expanded in the convenient HO basis functions. 
Therefore, in order to
exploit all the power of the HO algebra when manipulating the intrinsic operators, it
is pertinent to set the working $r_0$ - value equal to the respective ``optimum" value of the
oscillator parameter. 

In order to realize all the general results, obtained above for the intrinsic
density matrices and related distributions, 
the intrinsic wavefunctions  of the nuclear ground state have 
been constructed using the prescription by 
Ernst, Shakin and Thaler, that leads to the  so-called 
fixed-CM approximation. 
In this connection, we have demonstrated how one can unify 
the different approximate recipes of restoring the TI, if one starts with one of them, e.g., 
with the EST projection operator. 
As a specific example, analytic expressions for the 
intrinsic OBMD $\eta(\vec{p})$ and TBMD $\eta^{[2]}(\vec{p},\vec{k})$ 
of the \nuc{4}{He} nucleus have been derived 
within the context of the independent particle shell model, using harmonic-oscillator 
wavefunctions. When CM corrections are taken into account, 
the OBMD and TBMD are 
simultaneously 
shrunk with respect to the nontranslationally invariant counterparts. 
In addition, the CM correlation introduces in the case of the TBMD a dependence on the 
angle between $\vec{p}$ and $\vec{k}$. 
A shift of its peak for $p\neq 0$ 
in favor of opposite momenta and 
significant deviations for large values 
of $p$ and $k$ and angles close to 180$^{\circ}$ are 
observed. The above calculation is relevant 
to the current experimental study of two-nucleon 
knock out off {He} isotopes. For instance, we see similar 
behavior as the one found in the recent experimental study of the TBMD in 
\nuc{3}{He}~\cite{Nig04}. 

Of course, when increasing the momenta transferred to a residual system (in particular,
under kinematic conditions where they are getting comparable with the nucleon mass) the
corrections to TI breaking should be considered along with incorporating relativistic
effects such as the Lorentz contraction of nuclear WF's and a specific velocity-dependence
of nuclear forces. The latter is typical of relativistic
one-particle-exchange models (see survey \cite{ShSh01} and refs. therein), where, for instance, the
nucleon-nucleon
quasipotential, including recoil effects, is essentially nonlocal and prevents, in
contrast
to the nonrelativistic case, the separation of the CM motion of the relativistic system (nucleus)
from its internal motion. In other words, the corresponding four-momentum eigenstates
cannot be factorized as a product of independent CM and intrinsic components. In this
context, note a very instructive work \cite{HaG92}. Taking into account this distinctive feature of
relativistic quantum mechanics, it is difficult a priori to say to what extent the approach
developed here could be helpful in a covariant description of composite systems.
Nevertheless,
an encouraging example is found in Ref.~\cite{LTW98}, where the fixed-CM approximation combined with an
appropriate Lorentz contraction has been used for calculations of the nucleon
electromagnetic FF's in the cloudy bag model with CM and recoil corrections. 

Using the techniques presented in this paper 
one could calculate the intrinsic TBMD 
of other than \nuc{4}{He} $Z=N$, $\ell-$closed nuclei 
within the context of the harmonic oscillator 
model. In addition, within the present framework one could investigate the effects of 
short-range correlations on the intrinsic TBMD of 
\nuc{4}{He} and other $\ell-$closed nuclei, 
which are expected to be sizable for $p, k \geq p_F$, 
by introducing Jastrow-type correlations. 
Also, other intrinsic two-body quantities could be evaluated within the above 
general formalism (including other non-relativistic systems). 
Finally, the approach developed here may be helpful 
when evaluating the intrinsic one-body, 
two-body and more complicated density matrices in the HOM 
and in other independent-particle models 
(e.g.,  with 
single-particle wavefunctions of a potential well with finite depth, 
as it was shown in Ref.~\cite{KoS1977}).


\begin{thebibliography}{10}

\bibitem{Low55} 
P.O.~L{\"o}wdin, 
{Phys. Rev.} {\bf 97}, 1474 (1955).

\bibitem{BF90} 
O.~Benhar and A.~Fabrocini (Eds), 
{\em  Proceedings of the Workshop on Two-Nucleon Emission Reactions}, 
Elba International Physics Center, Italy, 1989, ETS Edition, Pisa 1990.

\bibitem{BGG03} 
A.~Braghieri, C.~Giusti and P.~Grabmayr (Eds), 
{\em Proceedings of the 6th Workshop on Electromagnetically Induced Two-Hadron Emission}, 
Pavia, Italy September 24-27, 2003.

\bibitem{Wal03} 
T.~Walcher, Prog. Part. Nucl. Phys. {\bf 50}, 503 (2003).

\bibitem{Nig04} 
R.A.~Niyazov {\em et al.}, Phys. Rev. Lett. {\bf 92}, 052303 (2004).

\bibitem{Sta04} 
A.V.~Stavinsky {\em et al.} (CLAS Collaboration), Phys. Rev. Lett. {\bf 93}, 192301 (2004).

\bibitem{PMK2000} 
P.~Papakonstantinou, E.~Mavrommatis and T.S.~Kosmas, Nucl. Phys. A {\bf 673}, 171 (2000). 

\bibitem{MPM2099} 
Ch.~Moustakidis, P.~Papakonstantinou and E.~Mavrommatis, in submission.

\bibitem{OrS95} 
G.~Orlandini and L.~Sarra, 
{\em Proceedings of the 2nd Workshop on Electromagnetically Induced Two-Nucleon Emission}, 
Gent, May 17-20, 1995, p.1. 

\bibitem{DKA2000} 
S.S.~Dimitrova, D.N.~Kadrev, A.N.~Antonov and M.V.~Stoitsov, Eur. Phys. J. A {\bf 7}, 335(2000).

\bibitem{PMK2003}
P.~Papakonstantinou, E.~Mavrommatis and T.S.~Kosmas,
\newblock {Nucl. Phys. A {\bf 713}}, 81 (2003). 

\bibitem{ES55}
J.P.~Elliot and T.H.~Skyrme,
\newblock {Proc. Roy. Soc. A {\bf 232}}, 561 (1955). 

\bibitem{DdF74} 
A.E.L.~Dieperink and T.~de Forest, Phys. Rev. C {\bf 10}, 543 (1974). 

\bibitem{dF80} 
T.~de Forest, Phys. Rev. C {\bf 22}, 2222 (1980).

\bibitem{TB58}
L.J.~Tassie and C.F.~Barker,
\newblock {Phys. Rev. {\bf 111}}, 940 (1958). 

\bibitem{GaS1957}
S.~Gartenhaus and C.~Schwartz,
\newblock {Phys. Rev. {\bf 108}}, 482 (1957). 

\bibitem{PeY1957}
R.~Peierls and J.~Yoccoz,
\newblock {Proc. Phys. Soc. A {\bf 70}}, 381 (1957). 

\bibitem{EST1973}
D.J.~Ernst, C.M.~Shakin, and R.M.~Thaler,
\newblock {Phys. Rev. C {\bf 7}}, 925 (1973).

\bibitem{EST1973b}
D.J.~Ernst, C.M.~Shakin, and R.M.~Thaler,
{\em ibid} 1340.

\bibitem{Vin1973}  
C.M.~Vincent, {Phys. Rev. C {\bf C}}, 929 (1973). 
 
\bibitem{ShG1974} 
A.V.~Shebeko and N.N.~Goncharov, Sov. J. Nucl. Phys. {\bf 18}, 532 (1974).

\bibitem{DOS1975}
S.~Dementiji, V.~Ogurtzov, A.~Shebeko and N.G.~Afanasiev,
\newblock {Sov. J. Nucl. Phys. {\bf 22}}, 6 (1976). 

\bibitem{Fri1971}
J.L.~Friar,
\newblock {Nucl. Phys. A {\bf 173}}, 257 (1971). 


\bibitem{ScG1990a}
K.W. Schmid and F.~Gr{\"u}mmer, 
\newblock {Z. Phys. A {\bf 336}}, 5 (1990). 

\bibitem{ScG1990}
K.W. Schmid and F.~Gr{\"u}mmer, 
\newblock {Z. Phys. A {\bf 337}}, 267 (1990). 

\bibitem{Schmid} 
K.W.~Schmid, 
Eur. Phys. J. A {\bf 12}, 29 (2001); 
{\em ibid} {\bf 13}, 319 (2002); 
{\em ibid} {\bf 14}, 413 (2002); 
{\em ibid} {\bf 16}, 475 (2003). 

\bibitem{RGS2004} 
R.R.~Rodriguez-Guzman and K.W.~Schmid, 
Eur. Phys. J. A {\bf 19}, 45 (2004); 
{\em ibid} {\bf 19}, 61 (2004). 

%\bibitem{Sch2001}
%K.W. Schmid.
%\newblock {\em Eur. Phys. J. {\bf A12}}, (2001).
%\newblock 29.
%
%\bibitem{Sch2002a}
%K.W. Schmid.
%\newblock {\em Eur. Phys. J. {\bf A13}}, (2002).
%\newblock 319.
%
%\bibitem{Sch2002b}
%K.W. Schmid.
%\newblock {\em Eur. Phys. J. {\bf A14}}, (2002).
%\newblock 413.

\bibitem{MiH99}
B.~Mihaila and J.H.~Heisenberg, Phys. Rev. C {\bf 60}, 054303 (1999).

\bibitem{GKS2002}
M.~Grypeos, C.~Koutroulos, A.~Shebeko and K.~Ypsilantis,
\newblock {Part. Nucl. Lett. {\bf 2}}, 111 (2002).  

\bibitem{VNW1998}
D.~Van Neck and M.~Waroquier,
\newblock {Phys. Rev. C {\bf 58}}, 3359 (1998). 

\bibitem{Pap2004}
P.~Papakonstantinou,
\newblock {\em {\rm PhD Thesis}}, University of Athens, 2004.

\bibitem{DSB1982}
M.~Dal Ri, S.~Stringari and O.~Bohigas, 
\newblock {Nucl. Phys. A {\bf 376}}, 81 (1982).

\bibitem{GW64}
M.L. Goldberger and K.M. Watson,
\newblock {\em Collision theory},
\newblock John Wiley and Sons, 1964.

\bibitem{BoM69} 
A.~Bohr and B.R.~Mottelson, 
\newblock {\em Nucleal Structure I}, 
A.~Benjamin, New York, 1969. 

\bibitem{NS69}
V.~Neudachin and Yu. Smirnov, 
\newblock {\em Nucleon clusters in light nuclei}.
\newblock {``Nauka"}: Moscow, 1964.

\bibitem{MYS67}
N.H. March, W.H. Young and S.~Sampanthar,
\newblock {\em The many-body problem in quantum mechanics}.
\newblock {Cambridge University Press, London}, 1967.

\bibitem{AHP88}
A.~Antonov, P.E. Hodgson and I.Zh. Petkov,
\newblock {\em Nucleon momentum and density distributions in nuclei}.
\newblock {Clarendon Press, Oxford}, 1988.

\bibitem{AHP93}
A.~Antonov, P.E. Hodgson and I.Zh. Petkov,
\newblock {\em Nucleon correlations in nuclei}.
\newblock {Springer-Verlag, Berlin-Heidelberg-New York}, 1993.

\bibitem{KoS1985}
A.Yu.~Korchin and A.V.~Shebeko,
\newblock {Z. Phys. A {\bf 321}}, 687 (1985) 
and refs. therein.

\bibitem{CPS84} 
C.~Ciofi degli Atti, E.~Pace and G.~Salme, 
Phys. Lett. {\bf 141B}, 14 (1984).  

\bibitem{KoS1977}
A.Yu.~Korchin and A.V.~Shebeko,
\newblock {Ukr. J. Phys. {\bf 22}}, 1646 (1977); 
arXiv: nucl-th/0601014.

\bibitem{PeT1962} 
R.E.~Peierls and D.J.~Thouless, 
\newblock {Nucl. Phys. {\bf 38}}, 154 (1962). 

\bibitem{KoV1992}
T.S. Kosmas and J.D. Vergados,
\newblock {Nucl. Phys. A {\bf 536}}, 72 (1992).

\bibitem{CdA1980} 
C.~Ciofi degli Atti, 
{Prog. Part. Nucl. Phys. {\bf 3}}, 163 (1980). 

\bibitem{She00} 
A.V.~Shebeko, {\em ``Lectures on Selected Topics 
of Nuclear Theory"}, Aristotle University of Thessaloniki, 
2000.  

\bibitem{DE88} 
{S.V.~Dementij et~al., {J. Phys. Soc. (Jap.)} {\bf 57}, 2988  
(1988). } 

\bibitem{VJV87}
H.~{de Vries}, C.W. {de Jager} and C.~{de Vries},
\newblock {At. Dat. Nucl. Dat. Tables {\bf 36}}, 495 (1987).

\bibitem{ShSh01} 
A.V.~Shebeko and M.I.~Shirokov, 
Prog. Part. Nucl. Phys. 44, 75 (2000); 
Phys. Part. Nucl. {\bf 32}, 31 (2001). 

\bibitem{HaG92} 
B.~Hamme and W.~Gloeckle, Few-Body Syst. {\bf 13}, 1 (1992). 

\bibitem{LTW98}
D.H.~Lu, A.W.~Thomas and A.G.~Williams, 
Phys. Rev. {\bf C57}, 2628 (1998).  

\end{thebibliography}
\end{document}